\documentclass[jcph]{apjrnl}

\usepackage{graphicx}
\begin{document}

\newcommand{\A}{{\cal A}}
\newcommand{\K}{{\cal K}}
\newcommand{\R}{{\mathbb R}}
\newcommand{\C}{{\mathbb C}}
\newcommand{\M}{{\mathbf M}}
\newcommand{\m}{{\mathbf m}}
\newcommand{\h}{{\mathbf h}}
\renewcommand{\H}{{\mathbf H}}
\newcommand{\pl}[2]{\frac{\partial#1}{\partial#2}}
\newcommand{\dl}[2]{\frac{\delta#1}{\delta#2}}
\newcommand{\p}{\partial}
\newcommand{\og}{\omega}
\newcommand{\Og}{\Omega}
\newcommand{\fl}[2]{\frac{#1}{#2}}
\newcommand{\dt}{\delta}
\newcommand{\td}{\tilde}
\newcommand{\tm}{\times}
\newcommand{\sm}{\setminus}
\newcommand{\nn}{\nonumber}
\newcommand{ \ap}{\alpha}
\newcommand{\bt}{\beta}
\newcommand{\ld}{\lambda}
\newcommand{\Gm}{\Gamma}
\newcommand{\gm}{\gamma}
\newcommand{\vp}{\varphi}
\newcommand{\tht}{\theta}
\newcommand{\ift}{\infty}
\newcommand{\vep}{\varepsilon}
\newcommand{\ep}{\epsilon}
\newcommand{\kp}{\kappa}
\newcommand{\Dt}{\Delta}
\newcommand{\Sg}{\Sigma}
\newcommand{\fa}{\forall}
\newcommand{\sg}{\sigma}
\newcommand{\ept}{\emptyset}
\newcommand{\btd}{\nabla}
\newcommand{\btu}{\bigtriangleup}
\newcommand{\tg}{\triangle}
\newcommand{\Th}{{\cal T} ^h}
\newcommand{\ged}{\qquad \Box}
\newcommand{\wt}{\widetilde}
\newcommand{\ol}{\overline}
\renewcommand{\theequation}{\arabic{section}.\arabic{equation}}
\newcommand{\be}{\begin{equation}}
\newcommand{\ee}{\end{equation}}
\newcommand{\ba}{\begin{array}}
\newcommand{\ea}{\end{array}}
\newcommand{\bea}{\begin{eqnarray}}
\newcommand{\eea}{\end{eqnarray}}
\newcommand{\beas}{\begin{eqnarray*}}
\newcommand{\eeas}{\end{eqnarray*}}
\newcommand{\dpm}{\displaystyle}

\title{\bf
 A Dynamic Atomistic-Continuum Method for the Simulation
of Crystalline Materials
\footnote{Abbreviated Title: Atomistic-Continuum Method for Crystals}}

\author{
Weinan E\ref{first}\ref{second} and Zhongyi Huang\ref{first}\ref{third} \bigskip \\
\additem[first]{Department of Mathematics and PACM, Princeton University, Princeton, NJ 08544 \bigskip}
\additem[second]{School of Mathematics, Peking University, Beijing, 100866, China } \bigskip 
\additem[third]{Department of Mathematical Sciences, Tsinghua University, Beijing, 100084, China} \bigskip
\\E-mail: weinan@princeton.edu, zhongyih@princeton.edu }

\date{November 30, 2001}

\maketitle

\begin{abstract}
We present a coupled atomistic-continuum method for the modeling of  
defects and interface dynamics of crystalline materials. The method
uses atomistic models such as molecular dynamics near defects and 
interfaces, and continuum models away from defects and interfaces.
We propose a new class of matching conditions between the atomistic
and continuum regions. These conditions ensure the accurate passage 
of large scale information between the atomistic and continuum regions 
and at the same time minimize the reflection of phonons at the 
atomistic-continuum interface. They can be made adaptive if we choose 
appropriate weight functions. We present applications to dislocation 
dynamics, friction between two-dimensional crystal surfaces and fracture 
dynamics. We compare results of the coupled method and the detailed 
atomistic model.
\end{abstract}

\newpage
\tableofcontents
\newpage

\section{\bf Introduction}
Traditionally two apparently separate approaches have been used to
model a continuous medium. The first is the continuum theory, in the 
form of partial differential equations describing the conservation laws 
and constitutive relations. This approach has been impressively successful 
in a number of areas such as solid and fluid mechanics. It is very efficient,
simple and often involves very few material parameters. But it  becomes 
inaccurate for problems in which the detailed atomistic processes affect
the macroscopic behavior of the medium, or when the scale of the medium
is small enough that the continuum approximation becomes questionable. 
Such  situations are often found in studies of properties and defects of 
micro- or nano- systems and devices. The second approach is atomistic, 
aiming at finding the detailed behavior of each individual atom using 
molecular dynamics or quantum mechanics. This approach can in principle 
accurately model the underlying physical processes. But it is often 
times prohibitively expensive.

Recently an alternative approach has been explored that couples the
atomistic and continuum approaches 
\cite{Tadmor1,Tadmor2,TimK1,TimK2,Rudd1,Rudd2,Cai}. 
The main idea is to use atomistic modeling at places where the displacement
field varies on an atomic scale, and the continuum approach elsewhere.
The most successful and best-known implementation is the quasi-continuum 
method \cite{Tadmor1,Tadmor2} which combines the adaptive finite element 
procedure with an atomistic evaluation of the potential energy of the system.
This method has been applied to a number of examples
\cite{Tadmor3, Tadmor4, Tadmor5}, 
and interesting details were learned about the structure of crystal defects.

Extension of the quasi-continuum method to dynamic problems has not been 
straightforward \cite{Rudd1,Rudd2,Cai}. The main difficulty lies in the 
proper matching between the atomistic and continuum regions. Since the 
details of lattice vibrations, the phonons, which are an intrinsic part 
of the atomistic model, cannot be represented at the continuum level,
conditions must be met that the phonons are not reflected at the 
atomistic-continuum interface. Since the atomistic region is expected to be
a very small part of the computational domain, violation of this condition
quickly leads to local heating of the atomistic region and destroys the 
simulation. In addition, the matching between the atomistic-continuum 
interface has to be such that large scale information is accurately 
transmitted in both directions.

The main purpose of the present paper is to introduce a new class of matching
conditions between atomistic and continuum regions. These matching conditions
have the property that they allow accurate passage of large scale (scales 
that are represented by the continuum model) information between the atomistic 
and continuum regions and no reflection of phonon energy to the atomistic 
region. These conditions can also be used in pure molecular dynamics
simulations as the border conditions to ensure no reflection of phonons
at the boundary of the simulation. As applications, we use our method to 
study the dynamics of dislocations in the Frenkel-Kontorova model, friction 
between crystal surfaces and crack propagation.

\section{\bf Continuum Approximation of Atomistic Models}
\setcounter{equation}{0}
As a first step toward constructing a coupled atomistic-continuum method,
we discuss briefly how continuum equations are obtained from atomistic models.

\subsection{1D Frenkel-Kontorova Model --- the Klein-Gordan Equation}
We first consider a simple problem, the Frenkel-Kontorova Model. This is 
a one-dimensional chain of particles in a periodic potential,
coupled by springs. We will take the potential to be:
\begin{equation}
\label{FK_potential}
U(x)=\frac{1}{2} {\cal K} (x-a\, \mbox{int}(x/a))^2
\end{equation}
Here $a$ is the equilibrium distance between neighboring particles, 
$\mbox{int}(x/a)$ is the integer part of $x/a$. Denote by $x_n$ the position
of the $n-$th particle, the dynamic equation for the particles is given by
\begin{equation}
\label{FK_force}
m\ddot{x}_n=k[x_{n+1}-x_n-a]+k[x_{n-1}-x_n+a]-U'(x_n) +f.
\end{equation}
where $f$ is the applied force.
%In the following, the spring constant $k$ is set to be 1, and so is the
%mass of the particles. 

One interesting aspect of the Frenkel-Kontorova  model is the possibility
of having dislocations in the system, which corresponds to vacant
or doubly occupied potential wells.
In the absence of dislocations, the equilibrium positions of the
particles are given by $x_j= ja$. In general, we 
let $x_j=a(j+u_j)$ and $\wt{f}=f/a$. 
$u$ is then the displacement field. A dislocation corresponds  to a kink
in $u$.
Far from the dislocations,
we can assume $|u_j - [u_j]| \ll 1$ where $[u]$ is the integer part
of $u$. Then we get, assuming $[u_j] = 0$,
\begin{equation}
\label{eq:FK_discrete}
m\ddot{u}_j=k[u_{j+1}-2u_j+u_{j-1}]-{\cal K} u_j +\wt{f}.
\end{equation}
Let $\tau=t\sqrt{m/(ka^2)}$, $\bar{\cal K}={\cal K}/(ka^{2})$, and 
$\bar{f}=\wt{f}/(k a^2)$, we obtain:
\begin{equation}
\label{eq:FK_discrete1}
\pl{^2{u}_j}{\tau^2}=u_{j+1}-2u_j+u_{j-1}-\bar{\cal K} u_j +\bar{f}.
\end{equation}
Taking the limit as $a \rightarrow 0$, we obtain 
the continuum limit equation for the displacement field $u$,
\be \label{FK_continuum}
\pl{^2 u}{\tau^2}=\pl{^2u}{x^2}-\bar{\cal K} u +\bar{f}.
\ee
This is simply the Klein-Gordan equation.

\subsection{2D Triangular Lattice --- Isotropic Elasticity}

Now we consider the triangular lattice model. We assume that nearest neighbor
atoms interact via central forces whose potential is given by
$\Phi(r^2)$ where $r$ is the distance between the atoms 
(see Figure \ref{tri_lattice}). From Newton's law, we have
\be \label{tri_motion}
m \ddot{\bf r}_{\bf 0}=-\sum_{\bf j}\btd_{\bf r_{0,j}}\Phi(|{\bf r_ {0,j}}|^2),
\ee
where $m$ is the mass of the atoms, ${\bf r_j}$ is the position
of the ${\bf j}$-th atom (${\bf j}=(j_1,j_2)$), $\bf r_{0,j}=r_0-r_j$. 
Let $\{\bf R_j\}$ be the equilibrium positions of the atoms. The lattice 
constant $a$ satisfies the equilibrium condition $\Phi'(a^2)=0$. Let 
$\{\bf u_j\}$ be the displacement vectors, ${\bf u_j = r_j - R_j}$.
Taylor expanding and omitting nonlinear terms in $\bf u$, we get
\bea
m\ddot{\bf u}_{\bf 0} &=& - \sum_{\bf j}\pl{}{|{\bf r_{0,j}}|^2} 
\Phi(|{\bf r_{0,j}}|^2) \pl{|{\bf r_{0,j}}|^2}{\bf r_{0, j}} \nn \\
&=& -2\sum_{\bf j} \Phi'(|{\bf r_{0, j}}|^2) {\bf r_{0, j}} \nn \\
&=& -2\sum_{\bf j} [\Phi'(a^2)+2\Phi''(a^2){\bf R_j}\cdot({\bf u_j-u_0})] 
[-{\bf R_j+ u_0- u_j}] \nn \\
&=& 4\Phi''(a^2) \sum_{\bf j} ({\bf R_j \otimes R_j}) ({\bf u_j-u_0}).
\label{eq:2d_tri_discrete}
\eea
Take the example of a Lennard-Jones potential:
\be \label{l_j_pot}
\Phi(r)=\ep_0\left(\fl{1}{(r/a_0)^{12}}-\fl{1}{(r/a_0)^6}\right),
\ee
then the lattice constant is equal to $a=2^{1/6}a_0$ under the assumption of
nearest neighbor interaction. In this case, equation (\ref{eq:2d_tri_discrete})
becomes
\bea
\fl{m}{\ep_0}\ddot{{\bf u}}_{\bf 0} &=& 
\fl{18}{a^2} \left\{ \left(\ba{cc}1 & 0 \\ 0& 0 \ea \right)
({\bf u}_{1,0}-2{\bf u}_{0,0}+{\bf u}_{-1,0})+ 
\left(\ba{cc} \fl{1}{4} & \fl{\sqrt{3}}{4} \\ \fl{\sqrt{3}}{4} & \fl{1}{4}\ea
\right) ({\bf u}_{0,1}-2{\bf u}_{0,0}+{\bf u}_{0,-1}) \right. \nn \\ 
&& \left. \quad
+\left(\ba{cc} \fl{1}{4} & -\fl{\sqrt{3}}{4}\\ -\fl{\sqrt{3}}{4} & \fl{1}{4}\ea
\right) ({\bf u}_{-1,1}-2{\bf u}_{0,0}+{\bf u}_{1,-1}) \right\}.
\label{eq:2d_tri_discrete_lj}
\eea
Let $\tau=t\sqrt{m/\ep_0}$. The continuum limit (as $a \to 0$) of the above
equations is
\bea
\pl{^2 {\bf u}}{\tau^2}&=&\left(\ba{cc} 81/4 & 0 \\ 0 & 27/4 \ea\right)\pl{^2{\bf u}}{x^2}
+\left(\ba{cc} 27/4 & 0 \\ 0 & 81/4 \ea \right) \pl{^2{\bf u}}{y^2}
+\left(\ba{cc} 0 & 54/4 \\ 54/4 & 0 \ea \right) \fl{\p^2{\bf u}}{\p x\p y} \nn \\
&=& (\ld+\mu) \btd(\btd\cdot {\bf u}) + \mu \btu {\bf u},
\eea
where $\ld=\mu=\fl{27}{4}$. This is the equation for isotropic elasticity.

\subsection{Slepyan Model of Fracture}
Here we give the one-dimensional and two-dimensional Slepyan models of 
fracture \cite{Marder}. In the 1D case, one can view it as a model for 
the atoms lying along a crack surface. Nearest neighbors are connected
by elastic springs, with spring constant $k$, and the atoms are tied to
the other side of the crack surface by similar springs,
which however snap when extended past some breaking point. The lines of 
atoms are being pulled apart by weak springs of spring constant $k/N$. 
These weak springs are meant schematically to represent $N$ vertical rows 
of atoms pulling in series. Let $\{u_{j,+},u_{j,-}\}$ be the displacement
of atoms on the top and bottom crack surfaces respectively. The equation 
which describes the upper line of mass points in this model is
\bea
m\ddot{u}_{j,+}=\left\{ \ba{l} 
k(u_{j+1,+}-2u_{j,+}+u_{j-1,+})  \\
+\fl{k}{N}(U_N-u_{j,+}) \\
+k(u_{j,-}-u_{j,+})\theta(2u_f-|u_{j,-}-u_{j,+}|) \\
-b\dot{u}_{j,+}
\ea \right.
\label{eq:1d_slepyan_fracture_model}
\eea
Here, the first term at the right hand side is elastic coupling to neighbors,
the second term is the driving force by displacing edges of the strip, the
third term is the bonding to atoms at the opposite side of the crack surface,
the last term is the dissipation,
$\theta$ is a step function, and the 
term containing it describes bonds which snap when their total extension
reaches a distance $2u_f$, where $u_f$ is a fracture distance. Assume the 
lattice constant is $a$. In the region far away from the fracture, we have
\be
m\ddot{u}_{j,+}=k(u_{j+1,+}-2u_{j,+}+u_{j-1,+})
+\fl{k}{N}(U_N-u_{j,+}) 
+k(u_{j,-}-u_{j,+})
-b\dot{u}_{j,+} 
\label{1d_slepyan_fracture_cont}
\ee
Dividing by $a^2$ and taking the limit as $a\to 0$, we obtain
\be \pl{^2u}{\tau^2}=\pl{^2u}{x^2}-\wt{b} \pl{u}{\tau}
\ee
with $\tau=t\sqrt{m/(ka^2)}$, $\wt{b}=b/(ka)$.

Now we consider a simple 2D model (see Figure \ref{sq_lattice}). A crack moves 
in a lattice strip composed of $2N$ rows of mass points. Assume that all the 
atoms are located at square lattice points if there is no exterior force on
them. All of the bonds between lattice points are brittle-elastic, behaving as
perfect linear springs until the instant they snap, from which point they 
exert no force. The displacement of each mass point is described by a single 
spatial coordinate $u_{i,j}$, which can be interpreted as the height of mass 
point $(i,j)$ into or out of the page. The index $i$ takes integer values, 
while $j=1/2-N,\cdots,-1/2,1/2,\cdots,N-1/2$. The model is described by the 
equation
\be \label{2d_slepyan_fracture_model}
m\ddot{u}_{i,j}=-b\dot{u}_{i,j}+ \sum_{\ba{c}{\rm nearest} \vspace{-4mm}\\ 
{\rm neighbors}\ (i',j')\ea} {\cal F} (u_{i',j'}-u_{i,j}),
\ee
with
\be \label{2d_slepyan_potential}
{\cal F} (r)=k r\theta(2u_f-|r|)
\ee
representing the brittle nature of the springs, $\tht$ the step function, 
and $b$ the coefficient of a small dissipative term. The boundary condition
which drives the motion of the crack is
\be \label{bc_2d_frac}
u_{i,\pm(N-1/2)}=\pm U_N.
\ee
Similarly, we can get the continuum limit of (\ref{2d_slepyan_fracture_model}):
\be \label{2d_slepyan_fracture_cont}
\pl{^2 u}{\tau^2}=\btu u - \wt{b}\pl{u}{\tau}.
\ee

\section{\bf Phonons}
\setcounter{equation}{0}
Among the most essential differences between the atomistic and continuum
behavior is the presence of phonons, the lattice vibrations, at the 
atomistic scale. In this section we will briefly review the spectrum of
the phonons. Let us first consider the simplest model: 1D discrete wave
equation (\ref{eq:FK_discrete}) with $k=1$, ${\cal K}=0$ and $f=0$. After 
discretization in time, we have
\begin{equation}
\label{eq:1D_discrete}
\frac{u^{n+1}_j-2u^n_j+u^{n-1}_j}{\Delta t^2}
=u^n_{j+1}-2u^n_j+u^n_{j-1}.
\end{equation}
where $u^n_j$ is the displacement of the $j$-th particle at time
$t=n\Delta t$.

The phonon spectrum for (\ref{eq:1D_discrete}) is obtained by looking for
solutions of the type $u^n_j=e^{i(n \omega \Delta t +j \xi )}$. This gives 
us the dispersion relation
\begin{equation}
\label{eq:1D_dispersion}
\frac1{\Delta t}\sin\frac{\omega\Delta t}2=\sin\fl{\xi}{2}.
\end{equation}
For the case when $\Dt t=0.01$, this dispersion relation is depicted in
Figure \ref{1d_dispersion}.
If ${\cal K} \ne 0$, we have
\begin{equation}
\label{eq:1D_dispersion_ne0}
\frac1{\Delta t}\sin\frac{\omega\Delta t}2=\sqrt{\sin^2\fl{\xi}{2}+{\cal K}/4}.
\end{equation}

Consider now the 2D triangular lattice described by (\ref{eq:2d_tri_discrete}).
Let us look for the solutions of the type
${\bf u}^n_{{\bf j}}=e^{i(\xi\cdot {\bf r_j} - n \omega \Delta t )} {\bf U}$ 
with ${\bf\xi}=(\xi_1,\xi_2)^T$. Substituting this expression into 
(\ref{eq:2d_tri_discrete}), we obtain
\bea
&&\left(\fl{\sin\frac{\omega\Delta t}{2}}{\Delta t/2}\right)^2 {\bf U} \nn \\
&=&\fl{18}{a^2}\left[\ba{cc}
4\sin^2\fl{\xi_1a}{2}+\sin^2\fl{\xi_1+\sqrt{3}\xi_2}{4}a
+\sin^2\fl{\xi_1-\sqrt{3}\xi_2}{4}a  & 
\sqrt{3}(\sin^2\fl{\xi_1+\sqrt{3}\xi_2}{4}a-\sin^2\fl{\xi_1-\sqrt{3}\xi_2}{4}a)\\ 
\sqrt{3}(\sin^2\fl{\xi_1+\sqrt{3}\xi_2}{4}a-\sin^2\fl{\xi_1-\sqrt{3}\xi_2}{4}a)&
3(\sin^2\fl{\xi_1+\sqrt{3}\xi_2}{4}a+\sin^2\fl{\xi_1-\sqrt{3}\xi_2}{4}a) \ea 
\right]{\bf U} \nn\\ 
&\equiv&{\bf A U}. \label{eq:2D_dispersion}
\eea
The eigenvalues of the matrix ${\bf A}$ are given by,
\be \label{eq:eignevalue_of_A}
\ld_{\pm}=\fl{36}{a^2}\left\{ \ap+\beta+\gm \pm 
\sqrt{\ap^2+\beta^2+\gm^2-\ap\beta-\ap\gm-\beta\gm}\right\},
\ee
where
\[ \ap=\sin^2\fl{\xi_1}{2}a, \qquad \beta=\sin^2\fl{\xi_1+\sqrt{3}\xi_2}{4}a, 
\qquad \gm=\sin^2\fl{\xi_1-\sqrt{3}\xi_2}{4}a.
\]
The dispersion relation now has two branches
\be
\omega_p(\xi_1,\xi_2)=\fl{2}{\Delta t}\arcsin(\fl{2}{\Delta t}\sqrt{\ld_+}),\quad
\omega_s(\xi_1,\xi_2)=\fl{2}{\Delta t}\arcsin(\fl{2}{\Delta t}\sqrt{\ld_-}),
\ee
where ``$p$'' and ``$s$'' stands for ``pressure'' and ``shear'' waves respectively.

\section{\bf Optimal Local Matching Conditions}
\setcounter{equation}{0}

We now come to the interface between the atomistic and continuum regions.
As we mentioned earlier, designing proper matching conditions at this
interface is a major challenge in such a coupled atomistic/continuum
approach.
The basic requirements for the matching conditions are the following:

(1). Reflection of phonons to the atomistic region should be minimal.
This is particularly crucial since the atomistic regions are typically
very small for the purpose  of computational efficiency, reflection of
phonon energy back to the atomistic region will trigger local heating
and melt the crystalline structure.

(2). Accurate exchange of large scale information  between the atomistic
and continuum regions.

The first requirement is reminiscent of the absorbing boundary conditions
required for the computation  of waves \cite{Clayton,Eng}. Indeed our work
draws much inspiration from that literature. There are some crucial
differences between the phonon problem considered here and the ones
studied in the literature on absorbing boundary conditions. The  most
obvious one is the fact that the electromagnetic or acoustic waves
are continuum objects modeled by partial differential equations, and
the associated absorbing boundary conditions often use small wavenumber
and/or frequency approximations, whereas the phonons are intrinsically
discrete with substantial energy distributed at high wavenumbers.

In the following we will give an example of a simple discrete wave
equation for which exact reflectionless boundary conditions can be
found. Such exact boundary conditions are highly nonlocal and therefore
not practical. But they give us guidelines on how approximate
boundary conditions should be constructed. We then present a method
that constructs optimal local matching conditions, given a predetermined
stencil.

\subsection{Exact Boundary Conditions for 1D Discrete Wave Equation}

Consider equation (\ref{eq:1D_discrete}). It is supposed to be solved for
all integer values of $j$. Now let us assume that we will truncate the
computational domain and only compute $u^n_j$ for $j\ge 0$. Assuming
there are no sources of waves coming from $j<0$, we still want to obtain
the same solution as if the computation is done for all $j$. At $j=0$, 
we will impose a new boundary condition to make sure that the phonons 
arriving from $j>0$ are not reflected back at $j=0$.

At $j=0$, we replace (\ref{eq:1D_discrete}) by
\begin{equation}
\label{eq:1D_abc}
u^n_{0}=\sum_{k,j\ge0}a_{k,j}u^{n-k}_j, \qquad a_{0,0} = 0.
\end{equation}
We would like to determine the coefficients $\{a_{k,j}\}$. For the simple 
problem at hand, it is possible to obtain analytical formulas of $\{a_{k,j}\}$
such that the imposition of (\ref{eq:1D_abc}) together with the solution of 
(\ref{eq:1D_discrete}) for $j>0$ reproduces exactly the solution of
(\ref{eq:1D_discrete}) if it was solved for all integer values of $j$, i.e. 
an exact reflectionless boundary condition can be found. 

First, let us consider the case of ${\cal K}=0$ and $f=0$. Let $\ld=\Dt t$ 
and let us look for solutions of the form:
\be \label{eq:1D_express}
u^n_j=z^n\xi^j, \qquad |\xi|\le 1. \ee
Substituting (\ref{eq:1D_express}) into (\ref{eq:1D_discrete}), we get
\be \label{eq:z_xi_relation}
\fl{1}{\ld^2}(z-2+\fl{1}{z}) = \xi-2+\fl{1}{\xi}.
\ee
This equation has two roots for $\xi$:
\be \label{eq:xi_of_z}
\xi_{1,2}=1+\fl{z^2-2z+1}{2\ld ^2z}\pm
\fl{1}{2\ld^2z}\sqrt{(z-1)^2[z^2+(4\ld^2-2)z+1]}.
\ee
Assume a boundary condition of the form
\be \label{eq:abc_1d_0}
u_0^{n+1}=2u_0^n-u_0^{n-1}+\ld^2(u_1^{n}-2u_0^n)+\sum_{k=1}^n s_k u_0^{n-k}.
\ee
Substituting (\ref{eq:1D_express}) into (\ref{eq:abc_1d_0}), we get
\[ z-2+\fl{1}{z}-\ld^2(\fl{1}{\xi}-2)=\sum_{k=1}^n s_k z^{-k}.\]
To find $s_k$, we have to find the Laurent expansion of the function 
on the left hand side. Let
\be \label{eq:hz_expansion}
H(z)=\sqrt{(z-1)^2[z^2+(4\ld^2-2)z+1]}.
\ee
Observe that $H(z)$ satisfies
\[ H'(z)=2(z-1)[z^2+(3\ld^2-2)z+1-\ld^2]/H(z).\]
Hence
\be \label{eq:hz_eqn}
H'(z)\cdot \{(z-1)[z^2+(4\ld^2-2)z+1]\}=2(z-1)[z^2+(3\ld^2-2)z+1-\ld^2] H(z).
\ee
Solving this equation by a Laurent series: 
$\dpm H(z)= \sum_{m\ge -2} \mu_m z^{-m}$,
we obtain a recursion relation $\mu_m$ for $m\ge 1$,
\be
(m+2)\mu_m=[1-2\ld^2-m(4\ld^2-3)]\mu_{m-1}
+[4-6\ld^2-m(3-4\ld^2)]\mu_{m-2}+(m-3)\mu_{m-3},
\label{eq:mu_recursion} \ee
and
\be \label{eq:mu_first}
\mu_{-2}=1, \quad \mu_{-1}=2\ld ^2-2, \quad \mu_0=1-2\ld^4.
\ee
Then from (\ref{eq:xi_of_z}) -- (\ref{eq:mu_first}), we have
\be \label{eq:1d_abc_final}
s_1=\ld^4, \qquad s_{k}=-\fl{\mu_{k-1}}{2}, \mbox{\ for \ } k\ge 2.
\ee
(\ref{eq:abc_1d_0}) is nonlocal and has memory effects. In order to see  
how fast the memory decays, let us assume $\mu_k \sim m^\ap$ when $m \gg 1$,
substituting into (\ref{eq:mu_recursion}), and equating the coefficients of
term order $m^{\ap}$, we get
\[ 2=(4\ld^2-3)(1+\ap)+(4-6\ld^2)+2(3-4\ld^2)(1+\ap)+(2\ld^2-2)-3(1+\ap).\]
This gives $\ap=-2$. The decay tendency of $\mu_k$ is shown in Figure 
\ref{fig_mu_decay}. Here $\ld=0.01$.

If $\K \ne 0$, we can proceed as before. But (\ref{eq:xi_of_z}) changes to
\be \label{eq:xi_of_z_kne0}
\xi_{1,2}=1+\fl{\cal K}{2}+\fl{z^2-2z+1}{2\ld ^2z}\pm \fl{1}{2\ld^2z}
\sqrt{[z^2+({\cal K}\ld^2-2)z+1][z^2+({\cal K}\ld^2+4\ld^2-2)z+1]}.
\ee
Assuming a boundary condition of the form
\be \label{eq:abc_1d_kne0}
u_0^{n+1}=2u_0^n-u_0^{n-1}+\ld^2[u_1^{n}-(2+{\cal K})u_0^n]
+\sum_{k=0}^n s_k u_0^{n-k},
\ee
Substituting (\ref{eq:1D_express}) into (\ref{eq:abc_1d_kne0}), we get
\[ z-2+\fl{1}{z}-\ld^2(\fl{1}{\xi}-2-\K)=\sum_{k=1}^n s_k z^{-k}.\]
To find $s_k$, we have to find the Laurent expansion of the function on
the left hand side. Let $g(\K,\ld)=\K\ld^2+2\ld^2-2$ and
\be \label{eq:hz_expansion_kne0}
H(z)=\sqrt{[z^2+({\cal K}\ld^2-2)z+1][z^2+({\cal K}\ld^2+4\ld^2-2)z+1]}.
\ee
Observe that $H(z)$ satisfies
\[ H'(z)=\{2z^3+3\,g(\K,\ld)z^2+[g^2(\K,\ld)+2-4\ld^4]z+g(\K,\ld)\}/H(z).\]
Hence
\bea 
&&H'(z)\cdot \{z^4+2\,g(\K,\ld)z^3+[g^2(\K,\ld)+2-4\ld^4]z^2+2\,g(\K,\ld)z+1\}
\nn \\ &=& H(z)\cdot\{2z^3+3\,g(\K,\ld)z^2+[g^2(\K,\ld)+2-4\ld^4]z+g(\K,\ld)\}.
\label{eq:hz_eqn_kne0} \eea
Solving this equation by a Laurent series: 
$\dpm H(z)=\sum_{m\ge -2}\mu_m z^{-m}$,
we obtain a recursion relation $\mu_m$ for $m\ge 2$,
\bea
(m+2)\mu_m&=&(2m+1)[2-\ld^2(\K+2)]\mu_{m-1}
+(1-m)\{2-4\ld^4+[2-\ld^2(\K+2)]^2\}\mu_{m-2} \nn \\
&& +(2-\K\ld^2-2\ld^2)(2m-5)\mu_{m-3} + (4-m)\mu_{m-4},
\label{eq:mu_recursion_kne0} \eea
and
\be \label{eq:mu_first_kne0}
\mu_{-2}=1, \quad \mu_{-1}=\K\ld^2+2\ld ^2-2, \quad \mu_0=1-2\ld^4, 
\quad \mu_1=2\ld^4(\K\ld^2+2\ld^2-2).
\ee
Then from (\ref{eq:xi_of_z_kne0}) -- (\ref{eq:mu_first_kne0}), we have
\be \label{eq:1d_abc_final_kne0}
s_0=-\ld^2{\K},\quad s_1=\ld^4, \quad s_{k}=-\fl{\mu_{k-1}}{2}, 
\ \mbox{for } k\ge 2.
\ee
In order to see how fast the memory decays, let us assume $\mu_k \sim m^\ap$ 
when $m \gg 1$, substituting into (\ref{eq:mu_recursion_kne0}), and equating 
the coefficients of term order $m^{\ap-1}$, we get
%\beas
%O(m^{\ap+1}): & 1=&{\cal K}({\cal K}+4) \ld^4 +1 \\
%O(m^{\ap}): & 2=&{\cal K}({\cal K}+4) \ld^4 (1+2\ap)+2 .
%\eeas 
\[ 0=\ap\{ g(\K,\ld)(2-\ap)-2\ap[g^2(\K,\ld)+2-4\ld^2]
-(6+9\ap)g(\K,\ld)-8-8\ap\}. \]
This gives $\ap=0$ or $\ap=-2/\left\{1+\fl{\K(\K+4)}{\K+2}\ld^2\right\}$.

These exact boundary conditions should be the same as the ones found
numerically in \cite{Cai}. It represents the exact Green's function 
for (\ref{eq:1D_discrete}) which is nonlocal. However, this procedure 
appears to be impractical for realistic models, particularly when the 
atomistic region moves with time which is the case that interests us. 
But such calculations can at least give us guidelines on how to proceed 
to construct approximately reflectionless boundary conditions.

\subsection{Optimal Local Matching Conditions for 1D Discrete Wave Equation}
A practical solution is to restrict (\ref{eq:1D_abc}) to a finite number of
terms and look for the
coefficients $\{a_{k,j}\}$ that minimize reflection. In order to do
this, let us look for solutions of
the type
\begin{equation}
\label{eq:1d_solu_wave}
u^n_j=e^{i(n\omega \Delta t+j\xi )}+R(\xi)e^{i(n\omega \Delta t -j\xi)}
\end{equation}
where $\omega$ is given by (\ref{eq:1D_dispersion}).
$R(\xi)$ is the reflection coefficient at wavenumber $\xi$. 
Inserting (\ref{eq:1d_solu_wave}) into (\ref{eq:1D_abc}), we obtain
\begin{equation}
\label{eq:1d_refl_coef}
R(\xi)=-\frac{\sum a_{k,j}e^{i(j\xi -k\omega\Delta t)}-1}{\sum
a_{k,j}e^{-i(j\xi +k\omega\Delta t)}-1}
\end{equation}
The optimal coefficients $\{a_{k,j}\}$ are obtained by
\begin{equation}
\label{eq:1d_min_refl}
\min\int^\pi_0 W(\xi) |R(\xi)|^2d\xi
\end{equation} 
subject to the constraint
\begin{equation}
\label{eq:1d_min_cons}
R(0)=0,R'(0)=0,\mbox{ etc.}
\end{equation}
Here $W(\xi)$ is a weight function, which is chosen to be $W(\xi) = 1$
in the examples below.

Condition (\ref{eq:1d_min_cons}) guarantees that large scale information is
transmitted accurately, whereas (\ref{eq:1d_min_refl}) guarantees that the
total amount of reflection is minimized. This procedure offers a lot of 
flexibility. For example, instead of $\int^\pi_0|R(\xi)|^2d\xi$, we can 
minimize the total reflection over certain carefully selected interval. Another
possibility is to choose the weight function to be the (empirically computed)
energy spectrum. The coefficients $\{a_{k,j}\}$ may then change in time to 
reflect the change of the nature of the small scales. In practice,
we found it preferable to use $\int^{\pi-\delta}_0|R(\xi)|^2d\xi$ with some
small $\delta$, instead of $\int^\pi_0|R(\xi)|^2d\xi$, in order to minimize
the influence of $\xi=\pi$ for which we always have $R(\pi)=1$.

Let us look at a few examples. If in (\ref{eq:1D_abc}) we only keep the terms 
involving $a_{1,0}$ and $a_{1,1}$, then imposing the condition $R(0)=0$ gives
\begin{equation}
\label{eq:1d_abc_first}
u^n_0 = (1 - \Delta t) u^{n-1}_0 +\Delta t u^{n-1}_1
\end{equation}
If instead we keep terms involving $a_{0,1}, a_{1,0}$ and $a_{1,1}$, we can 
then impose both $R(0)=0$ and $R'(0) = 0$. This gives us
\begin{equation}
\label{eq:1d_abc_second}
u_0^n = u_1^{n-1} + \frac{1-\Delta t}{1+\Delta t}(u_0^{n-1} - u_1^n)
\end{equation}
Conditions of the type (\ref{eq:1d_abc_first}) and (\ref{eq:1d_abc_second}) are
intimately related to the absorbing boundary conditions proposed and analyzed 
in \cite{Clayton,Eng} for the computation of waves. These conditions perform 
well for low wavenumbers but are less satisfactory at high wavenumbers.

To improve the performance at high wavenumbers let us consider a case that 
include terms with $k\le2,j\le3$ and minimize $\int^{\pi-\dt}_0|R(\xi)|^2d\xi$ 
(with $\delta=0.125\pi$) subject to the condition $R(0)=0$, the optimal 
coefficients can be easily found numerically and are given by
\begin{equation}
\label{eq:1d_abc_coef_23}
(a_{k,j})=\left(\begin{array}{ccc}
1.95264 & -7.4207\times10^{-2} & -1.4903\times10^{-2} \\ [.15in]
-0.95406 & 7.4904\times10^{-2} &
1.5621\times10^{-2}\end{array}\right)
\end{equation}
If instead we only include terms such that $k\le3,j\le2$, then
\begin{equation}
\label{eq:1d_abc_coef_32}
(a_{k,j})=\left(\begin{array}{cc}
2.9524 & 1.5150\times10^{-2} \\ [.15in]
-2.9065 & -3.0741\times10^{-2} \\ [.15in]
0.95406 & 1.5624\times10^{-2}\end{array}\right)
\end{equation}
The resulting reflection coefficients $R$ are displayed in Figure
\ref{reflect_coef}.

\subsection{Optimal Local Matching Conditions for Triangular Lattice}
The above procedure can be easily generalized. Let us take the triangular
lattice as an example, and the boundary to be the $x$-axis. Given a boundary
condition of the form:
\be \label{eq:2d_abc}
{\bf u}^{n+1}_0 = \sum_{l\le 1} \sum_{\bf j} {\bf A}^l_{\bf j} 
{\bf u}^{n+l}_{\bf j},
\ee
where ${\bf A}^l_{\bf j}$ are some $2\times 2$ matrices, and the summation
is done over a pre-selected stencil, we can find the reflection matrix
associated with this boundary condition. For that purpose,
we look for solutions of the form
\be \label{eq:2d_solution}
{\bf u}^n_{\bf j} = \sum_{\ap=I,O}\sum_{\beta=s,p} C^\ap_\beta 
e^{i(\xi^{\ap\beta} \cdot {\bf r}_{\bf j}-\omega t)} {\bf U}^\ap_\beta,
\ee
where $\ap=I,O$ correspond to ``incoming'' and ``outgoing'' waves 
respectively (see Figure \ref{fig_reflect}),
$\beta=s,p$ correspond to ``shear'' and ``pressure'' waves respectively.
Substituting into (\ref{eq:2D_dispersion}) we obtain a relation between
$(C^O_s, C^O_p)^T$ and $(C^I_s, C^I_p)^T$,
\be \label{eq:reflect_matrix}
{\bf M}_{\bf 0} \left( \ba{c}C^O_s \\ C^O_p \ea\right) 
= {\bf M}_I\left( \ba{c}C^I_s \\ C^I_p\ea\right),
\ee
where
\bea \label{eq:matrix_Out}
{\bf M_0}&=& e^{-i\og \Dt t} [U^O_s,\ U^O_p] - \sum_{l\le 1}\sum_{\bf j} 
{\bf A}_{\bf j}^l \left[ e^{i(\xi^{Os}\cdot{\bf r_j}-l\omega \Dt t)}U^O_s,
\  e^{i(\xi^{Op} \cdot{\bf r_j}-l\og \Dt t)}U^O_p\right], \\
\label{eq:matrix_In}
{\bf M}_I&=&-e^{-i\og \Dt t} [U^I_s,\ U^I_p] + \sum_{l\le 1}\sum_{\bf j} 
{\bf A}_{\bf j}^l \left[ e^{i(\xi^{Is}\cdot{\bf r_j}-l\omega \Dt t)}U^I_s,
\  e^{i(\xi^{Ip} \cdot{\bf r_j}-l\og \Dt t)}U^I_p\right].
\eea
In principle, we can solve the minimization problem
\be \label{eq:reflect_matrix_norm}
\mbox{min }\int W(\xi) \|{\bf M}_I ^{-1}\cdot {\bf M_0}(\xi) \|^2  d\xi
\ee
to find optimal $\{{\bf A_j}^l\}$, where the integration is over the Brillion
zone. But in practice, we find it much more convenient to restrict the
integration over a few selected low symmetry atomic planes. In the present
context, it amounts to choosing special incidences where the phonons energy
dominates.

First, let us consider the case of normal incidence $\tht=90^\circ$.
That means $\xi_1=0$. Then the matrix ${\bf A}$ in (\ref{eq:2D_dispersion}) 
becomes a diagonal matrix:
\beas
{\bf A}=\fl{18}{a^2}\left[ \ba{cc} 2\sin^2\fl{\sqrt{3}\xi_2 a}{4}& 0 \\
0 & 6\sin^2\fl{\sqrt{3}\xi_2 a}{4} \ea\right].
\eeas
with two eigenvalues and eigenvectors:
\beas
\ld_1= \fl{36}{a^2}\sin^2\fl{\sqrt{3}\xi_2 a}{4}, & {\bf U}_1=(1,0)^T, \\
\ld_2= \fl{108}{a^2}\sin^2\fl{\sqrt{3}\xi_2 a}{4}, & {\bf U}_2=(0,1)^T.
\eeas
Then dispersion relations are
\bea \label{eq:thteq90_eigv1}
\og_s\Dt t= 2\arcsin(\fl{3\Dt t}{a}\sin\fl{\sqrt{3}\xi_2 a}{4}), 
& {\bf U}_s=(1,0)^T, \\
\og_p\Dt t= 2\arcsin(\fl{3\sqrt{3}\Dt t}{a}\sin\fl{\sqrt{3}\xi_2 a}{4}), 
& {\bf U}_p=(0,1)^T.
\label{eq:thteq90_eigv2}
\eea
If we take the absorb boundary condition as in (\ref{eq:2d_abc}), the matrices
${\bf M_0}$ and ${\bf M_I}$ are
\bea \label{eq:matrix_O90}
{\bf M_0}&=& e^{-i\og \Dt t} I - \sum_{l\le 1}\sum_{\bf j} {\bf A}_{\bf j}^l 
\left( \ba{cc} e^{i(\xi_2^{Os}y_{\bf j}-l\omega \Dt t)} & 0 \\ 
0 & e^{i(\xi_2^{Op}y_{\bf j}-l\omega \Dt t)}\ea\right), \\
\label{eq:matrix_I90}
{\bf M}_I&=&-e^{-i\og \Dt t} I + \sum_{l\le 1}\sum_{\bf j} {\bf A}_{\bf j}^l
\left( \ba{cc} e^{i(\xi_2^{Is}y_{\bf j}-l\omega \Dt t)} & 0 \\ 
0 & e^{i(\xi_2^{Ip}y_{\bf j}-l\omega \Dt t)}\ea\right),
\eea
where $I$ is the $2\tm 2$ identity matrix. For consistency, we should require
that the low wavenumber waves be transmitted accurately. Imposing
(\ref{eq:reflect_matrix_norm}), we get
\bea \label{eq:2d_constrains90_1}
I&=&\sum_{l\le 1} \sum_{\bf j} {\bf A} _{\bf j}^l ,\\
0&=&\Dt t\ I + \sum_{l\le 1}\sum_{\bf j} {\bf A} _{\bf j}^l 
\left( \ba{cc} \fl{2\sqrt{3}a}{9}y_{\bf j}-l\Dt t & 0 \\ 
0 & \fl{2a}{9}y_{\bf j}-l\Dt t \ea\right)
\label{eq:2d_constrains90_2}\eea
If we minimize (\ref{eq:reflect_matrix_norm}) along normal incidence subject to
the constraints (\ref{eq:2d_constrains90_1}) and (\ref{eq:2d_constrains90_2}),
we obtain the desired matrices ${\bf A_j}^l$. For example, if we keep the terms 
with $l=0,1$ and ${\bf j}=(0,0)$, $(-1,0)$, $(-1,1)$, 
%(see Figure \ref{tri_lattice} and \ref{fig_reflect}), 
the optimal coefficient matrices are
\beas
{\bf A}^0_{(0,0)}&=& \left(\ba{cc}   
    0.947937634      & -0.423061769E-09 \\
   -0.411523005E-09  &  0.911511476 \ea \right)\\
{\bf A}^0_{(-1,0)}={\bf A}^0_{(-1,1)}&=& \left(\ba{cc}   
    0.500000011      &  0.604865049E-08 \\
    0.105260341E-07  &  0.499999996 \ea \right)\\
{\bf A}^1_{(-1,0)}={\bf A}^1_{(-1,1)}&=& \left(\ba{cc}   
   -0.473968784      & -0.603638049E-08 \\
   -0.102331855E-07  & -0.455755718 \ea \right)
\eeas

Next we consider the cases when both $\tht=60^\circ$ and $\tht=120^\circ$ are
taken into account. For $\tht=60^\circ$, we have $\xi_2=\sqrt{3}\xi_1$, and
\beas
{\bf A}=\fl{18}{a^2}\sin^2\fl{\xi_1 a}{2}\left[\ba{cc}9-4\sin^2\fl{\xi_1 a}{2}&
\sqrt{3}(3-4\sin^2\fl{\xi_1 a}{2})\\ 
\sqrt{3}(3-4\sin^2\fl{\xi_1 a}{2}) & 15-12\sin^2\fl{\xi_1 a}{2} \ea\right],
\eeas
with two eigenvalues and eigenvectors:
\beas \ba{rcll}
\ld_1&=& \fl{108}{a^2}\sin^2\fl{\xi_1 a}{2}, & {\bf U}_1=(\sqrt{3},-1)^T, \\
\ld_2&=& \fl{36}{a^2}\sin^2\fl{\xi_1 a}{2}(9-8\sin^2\fl{\xi_1 a}{2}), 
& {\bf U}_2=(1,\sqrt{3})^T. \ea
\eeas
The dispersion relations are
\bea \label{eq:thteq60_eigv1}
\og_s\Dt t&=&2\arcsin\left(\fl{3\sqrt{3}\Dt t}{a}\sin\fl{\xi^s_1 a}{2}\right),
\qquad \qquad \qquad {\bf U}_s=(\sqrt{3},-1)^T, \\
\og_p\Dt t&=&2\arcsin\left(\fl{3\Dt t}{a}\sin\fl{\xi^p_1 a}{2}
\sqrt{9-8\sin^2\fl{\xi^p_1 a}{2}}\right), \quad {\bf U}_p=(1,\sqrt{3})^T.
\label{eq:thteq60_eigv2}
\eea
The consistency constraints are 
\bea \label{eq:2d_constrains60_1}
I&=&\sum_{l\le 1} \sum_{{\bf j}} {\bf A}_{{\bf j}}^l ,\\
0&=&\Dt t\left( \ba{cc}\sqrt{3}&1\\-1&\sqrt{3} \ea\right) + 
\sum_{l\le 1}\sum_{\bf j} {\bf A}_{\bf j}^l \left( \ba{cc} 
\sqrt{3}(\xi^{Is}\cdot{\bf r_j}/\omega -l\Dt t) 
& \xi^{Ip}\cdot{\bf r_j}/\og -l\Dt t \\ 
-(\xi^{Is}\cdot{\bf r_j}/\og -l\Dt t) &
\sqrt{3} (\xi^{Ip}\cdot{\bf r_j}/\og -l\Dt t) \ea\right) \qquad\
\label{eq:2d_constrains60_2}\eea
for $\tht=60^\circ$, and
\bea \label{eq:2d_constrains120_1}
I&=&\sum_{l\le 1} \sum_{\bf j} {\bf A}_{\bf j}^l ,\\
0&=&\Dt t\left( \ba{cc}\sqrt{3}&-1\\1&\sqrt{3} \ea\right) + 
\sum_{l\le 1}\sum_{\bf j} {\bf A}_{\bf j}^l 
\left( \ba{cc} \sqrt{3}(\xi^{Is}\cdot{\bf r_j}/\og -l\Dt t) & 
-(\xi^{Ip}\cdot{\bf r_j}/\og -l\Dt t )
\\ \xi^{Is}\cdot{\bf r_j}/\og -l\Dt t &
\sqrt{3} (\xi^{Ip}\cdot{\bf r_j}/\og -l\Dt t) \ea\right) \qquad\
\label{eq:2d_constrains120_2}\eea
for $\tht=120^\circ$.
For example, if we keep the terms for $l=0,1$ and ${\bf j}=(0,0)$, $(-1,0)$,
$(-1,1)$, we have the optimal coefficient matrices
\beas
{\bf A}^0_{(0,0)}&=& \left(\ba{cc}   
   0.929252841     & -0.861918368E-09 \\
   0.355336047E-09 &  0.908823412 \ea \right)\\
{\bf A}^0_{(-1,0)}&=& \left(\ba{cc}   
   0.504255087     &  0.156041692E-01 \\
   0.161793547E-01 &  0.499201553 \ea \right)\\
{\bf A}^0_{(-1,1)}&=& \left(\ba{cc}
   0.504255087     & -0.156041692E-01 \\
  -0.161793547E-01 &  0.499201553 \ea \right)\\
{\bf A}^1_{(-1,0)}&=& \left(\ba{cc}   
  -0.468881506     &  0.128308044E-01 \\
   0.118075167E-01 & -0.453613259 \ea \right)\\
{\bf A}^1_{(-1,1)}&=& \left(\ba{cc}
  -0.468881506     & -0.128308044E-01 \\
  -0.118075167E-01 & -0.453613259 \ea \right)
\eeas
If all three angles $\theta = 60^\circ, 90^\circ, 120^\circ$ are used with equal
weight, then the optimal coefficient matrices are given by:
\beas
{\bf A}^0_{(0,0)}&=& \left(\ba{cc}   
    0.963685659E+00 &  0.522045701E-05 \\
    0.186532512E-05 &  0.911580620E+00
 \ea \right)\\
{\bf A}^0_{(-1,0)}&=& \left(\ba{cc}   
    0.190155146E+00 &  0.439544149E-02 \\
    0.132862553E-01 &  0.497292487E+00
 \ea \right)\\
{\bf A}^0_{(-1,1)}&=& \left(\ba{cc}
    0.190158427E+00 & -0.439289996E-02 \\
   -0.132859770E-01 &  0.497292916E+00
 \ea \right)\\
{\bf A}^1_{(-1,0)}&=& \left(\ba{cc}   
   -0.171943945E+00 &  0.439598834E-02 \\
    0.132858254E-01 & -0.459575584E+00
 \ea \right)\\
{\bf A}^1_{(-1,1)}&=& \left(\ba{cc}
   -0.171944818E+00 & -0.439293875E-02 \\
   -0.132856732E-01 & -0.453075576E+00
 \ea \right)
\eeas

\section{\bf Algorithms and Implementations}

The basic framework of our coupled continuum/atomistic method is
that of an adaptive mesh refinement method \cite{berger}. The computational
domain is covered by  a grid that resolves the macroscopic features of
problems, such as applied forces and boundary conditions. Regions near
atomistic defects such as dislocations, interfaces, cracks, impurities, etc
are detected using some error estimators. Molecular dynamics are
used in these regions to compute the location and momentum of each atom,
together with the averaged quantities at the macroscopic grid points.
Continuum equations are used elsewhere. At the interface between 
the two regions, matching conditions discussed in the last section are
used. Specifically, we decompose the velocity and displacement
fields into a large scale part and a small scale part. The large
scale part is evolved using the values at the macroscopic grid
points. In the atomistic regions, these are the averaged quantities.
The small scale part is computed using the reflectionless
boundary conditions discussed above.

One important aspect of this method is the error estimators that 
are used to distinguish atomistic and continuum regions. 
The senstivity of the error estimators determines the balance between
accuracy and efficiency. However, since there
has already been a lot of work done 
on this specific problem \cite{babuska,oden,rannacher}, we will not
pursue this question here further. We find it adequate in our work
to use a refinement indicator (rather than an error estimator) which
is given either by an estimate of the stress, or a weighted average
of the wavelet coefficients.

Further details of our method are explained through a series of 
examples.

\setcounter{equation}{0}
\subsection{Dislocation Dynamics in the Frenkel-Kontorova Model}
As the simplest model that encompasses most of the issues in a coupled 
atomistic/continuum simulation, we consider the Frenkel-Kontorova model
\begin{equation}
\label{eq:1d_fk_keq0}
\ddot{x}_j=x_{j+1}-2x_j+x_{j-1}-U'(x_j)+f
\end{equation}
where $U$ is a periodic function with period 1, $f$ is an external forcing.
The continuum limit of this equation is simply the Klein-Gordan equation
\begin{equation} \label{eq:1d_fk_cont}
u_{tt}=u_{xx}-Ku+f
\end{equation}
where $K=U''(0)$. We consider the case when there is a dislocation and study
its dynamics under a constant applied forcing. We use $U(x)=(x-[x])^2$ where
$[x]$ is the integer part of $x$. In this example we take 
(\ref{eq:1d_fk_keq0})
as our atomistic model, and (\ref{eq:1d_fk_cont}) as our continuum model.
For the coupled atomistic-continuum method, we use a standard second order 
finite difference method for (\ref{eq:1d_fk_cont}) in the region away from 
the dislocation, and we use (\ref{eq:1d_fk_keq0}) in the region around the
dislocation. However, we also place finite difference grid points in the
atomistic region. At these points, the values are obtained through averaging
the values from the atomistic model. At the interface between the atomistic
and continuum regions, we decompose the displacement into a large scale and
a small scale part. The large scale part is computed on the finite difference
grid, using (\ref{eq:1d_abc_coef_32}). The small scale part is computed using
the reflectionless boundary conditions described earlier. The interfacial
position between the MD and continuum regions is moved adaptively according 
to an analysis of the wavelet coefficients or  the local stress. The two
strategies lead to similar results. Care has to be exercised in order to 
restrict the size of the atomistic region. For example, when wavelet
coefficients are used in the criteria to move the atomistic region, we found
it more efficient to use the intermediate levels of the wavelet coefficients
rather than the finest level.

We first consider the case when a sharp transition is made between the
atomistic and continuum regions with a 1:16 ratio for the size of the grids.
Figure \ref{fig_ex_fk1} is a comparison of the displacement and velocity 
fields computed using the full atomistic  model and the coupled
atomistic/continuum model, with $f=0.04$. The atomistic region has 32 atoms.
The full atomistic simulation has $10^3$. Dislocation appears as a kink in 
the displacement field. Notice that at the atomistic/continuum interface, 
there is still substantial phonon energy which is then suppressed by the 
reflectionless boundary condition. No reflection of phonons back to the 
atomistic region is observed. In Figure \ref{dislocation_pos_04}, we compare
the positions of the dislocation as a function of time, computed using the
coupled method and the detailed molecular dynamics. Extremely good agreement
is observed.

We next consider a case with $f=0.02$, which alone is too weak to move
the dislocation, but to the left of the dislocation, we add a sinusoidal 
wave to the initial data. The dislocation moves as a consequence of the 
combined effect of the force and the interaction with the wave. Yet in this
case the same atomistic/continuum method predicts an incorrect position for
the dislocation, as shown in Figure \ref{fig_ex_fk2}. The discrepancy seems to
grow slowly in time (see Figure \ref{dislocation_pos_02}). Improving the 
matching conditions does not seem to lead to significant improvement.

The difference between this case and the case shown in Figure \ref{fig_ex_fk1}
is that there is substantially more energy at the intermediate scales. This
is clearly shown in the energy spectrum that we computed for the two cases
but it can also be seen in Figure \ref{fig_ex_fk2} where an appreciable amount
of small scale waves are present in front of the dislocation. 
Such intermediate
scales are suppressed in a method that uses a sharp transition between the
atomistic and continuum regions, unless we substantially increase the size 
of the atomistic region. We therefore consider the next alternative 
in which the atomistic/continuum transition is made gradually in a 1:2 or 1:4 
ratio between neighboring grids. The right column in Figure \ref{fig_ex_fk2} 
shows the results of such a method that uses a gradual 1:2 transition. 
We see that the correct dislocation position is now recovered.

\subsection{Friction between Flat and Rough Crystal Surfaces}
Our second example is the friction between crystal surfaces.
To model this process atomistically, we use standard molecular dynamics with
the Lennard-Jones potential \cite{Harrison,Robbins}. First, we consider the 
case in which the two crystals are separated by a horizontal atomically
flat interface. The
atoms in the bottom crystal are assumed to be much heavier 
(by a factor of 10) than the atoms on
top. To model the lack of chemical bonding between the atoms in the top
and bottom crystals, the interaction forces are reduced by a factor of
5 between atoms in the top and bottom crystals. 
A constant shear stress is applied near the top surface. 
We use the periodic boundary condition
in the $x$-direction. 

From a physical viewpoint, one interesting issue
here is how dissipation takes place. Physically the  kinetic energy of
the small scales appears as phonons which then convert into heat and
exit the system. A standard practice in modeling such a process is to add a
friction term to the molecular dynamics in order to control the temperature
of the system \cite{Harrison,Robbins}. In contrast, we ensure the proper 
dissipation of phonons to the environment by imposing the reflectionless 
boundary conditions for the phonons. 
The results presented below are computed using the last set
of coefficient matrices presented at the end of Section 4.

From Figure \ref{fig_flat_fric} we see that we indeed 
obtain a linear relation between
the mean displacement of the atoms in the top crystal 
as a function of time. 
The temperature of the system also saturates.
Also plotted in Figure \ref{fig_flat_fric} is the result of the mean
displacement computed using the combined atomistic/continuum method. Here
the continuum model is the linear elastic wave 
equation with Lame coefficients
computed from the Lennard-Jones potential. The agreement between the full
atomistic and the atomistic/continuum simulation is quite satisfactory.

Next, we study the friction between two rough crystal surfaces. The setup
is the same as before, except that the initial interface between 
the crystals
takes the form $y=f(x)$. The numerical results obtained are displayed
in Figure \ref{fig_rough_fric}.
In Figure \ref{friction_interface}, we plot the positions of the atoms in the top
and bottom crystals.  We see that gaps are created in the case of
rough interfaces.

In Figure \ref{vel_force}, we compare the force-velocity relations 
for both flat and rough interfaces. Again the agreement between 
the coupled method and the full atomistic method is quite good.

In the present problem, we used atomistic model in a narrow strip near the
interface, and continuum model away from the interface. An interesting
question is how wide the atomistic strip has to be. Clearly for the purpose
of computational efficiency, we want the atomistic strip to be as narrow as
possible. On the other hand, it has to be wide enough to provide an
accurate description inside the  boundary
layer where important atomistic processes can be relaxed. There are
two important atomistic processes in the present problem. The first is the
vibration of the atoms around their local equilibrium positions. 
The second is
the process of moving from one local equilibrium to the next, i.e. sliding
by one atomic distance. Clearly the second process works on longer time
scale.
This process has to be resolved by the atomistic layer. In Figure
\ref{fig_friction_1d}, we compare the atomic positions 
of a column of atoms which were initially vertical, i.e. they had the same
$x$-coordinates. From this picture one can also estimate the strain rate.
We can clearly see that if the atomistic layer does not resolve the phonons
generated by the second process, we get inaccurate results. 

\subsection{Crack Propagation}
Our third example is the Slepyan  model of fracture dynamics
(\ref{eq:1d_slepyan_fracture_model}). In our coupled atomistic/continuum 
method, we use full atomistic simulation (\ref{eq:1d_slepyan_fracture_model})
around the crack tip, and use (\ref{1d_slepyan_fracture_cont}) in the region
far away from the crack tip. For the continuum equation, we use the 
displacement boundary
condition $u_{\pm}=\pm U_N$  at the left boundary,
and stress boundary condition $\frac{\partial u}{\partial x} = 0$
at the right boundary.
Figure \ref{fig_1d_fracture}
is a comparison of the fracture surface computed using the 
full atomistic model
and the coupled atomistic/continuum method.

Next we apply our method to the 2D Mode III fracture dynamics on a
square lattice (\ref{2d_slepyan_fracture_model}). 
Same boundary conditions as in the 1D case are used for the continuum model.
For the matching conditions between the atomistic and continuum
regions, we used a stencil that consists of seven points: the values
of the three nearest grid points next to the boundary at the current
and previous time steps, plus the value at the boundary grid point at the
previous time step. The optimization is carried out using angles
$\theta = 45^\circ, 90^\circ, 135^\circ$.
Figure \ref{fig_2d_fracture} is a comparison 
of the fracture surface computed using the 
full atomistic model and the coupled 
atomistic/continuum method.
Comparisons of the positions of the fracture tip as a function of
time is given in Figure \ref{crack_tip}.
The results are quite satisfactory.
Finally in Figure \ref{shearwave}, we display the shear waves generated
as a result of the crack propagation, No reflection is seen.

\section{\bf Conclusion}
In conclusion, we presented a new strategy for the matching conditions
at the atomistic/continuum interface in  multiscale modeling of crystals.
The main idea is to choose the boundary condition by minimizing the reflection
of phonons along a few low symmetry atomic planes, subject to some accuracy
constraints at low wavenumbers. These conditions are adaptive if we choose 
the weight functions in (\ref{eq:1d_min_refl}) and 
(\ref{eq:reflect_matrix_norm}) to reflect the evolving nature of the small 
scales. They minimize the reflection of phonons and at the same time ensure 
accurate passage of large scale information. The coupled atomistic/continuum
method presented here is quite robust and works well at low temperature. At
finite temperature and when nonlinearity is important at large scales, a
new method has to be worked out. This work is in progress.

We thank Tim Kaxiras for suggesting the problem of friction
between rough interface.
This work is supported in part by NSF through a PECASE award and by ONR grant
N00014-01-1-0674.

\newpage
\begin{center}
{\bf \large List of Figures}
\end{center}
\medskip

\begin{figure} %1
\begin{center}
\resizebox{5.5in}{!}{\includegraphics{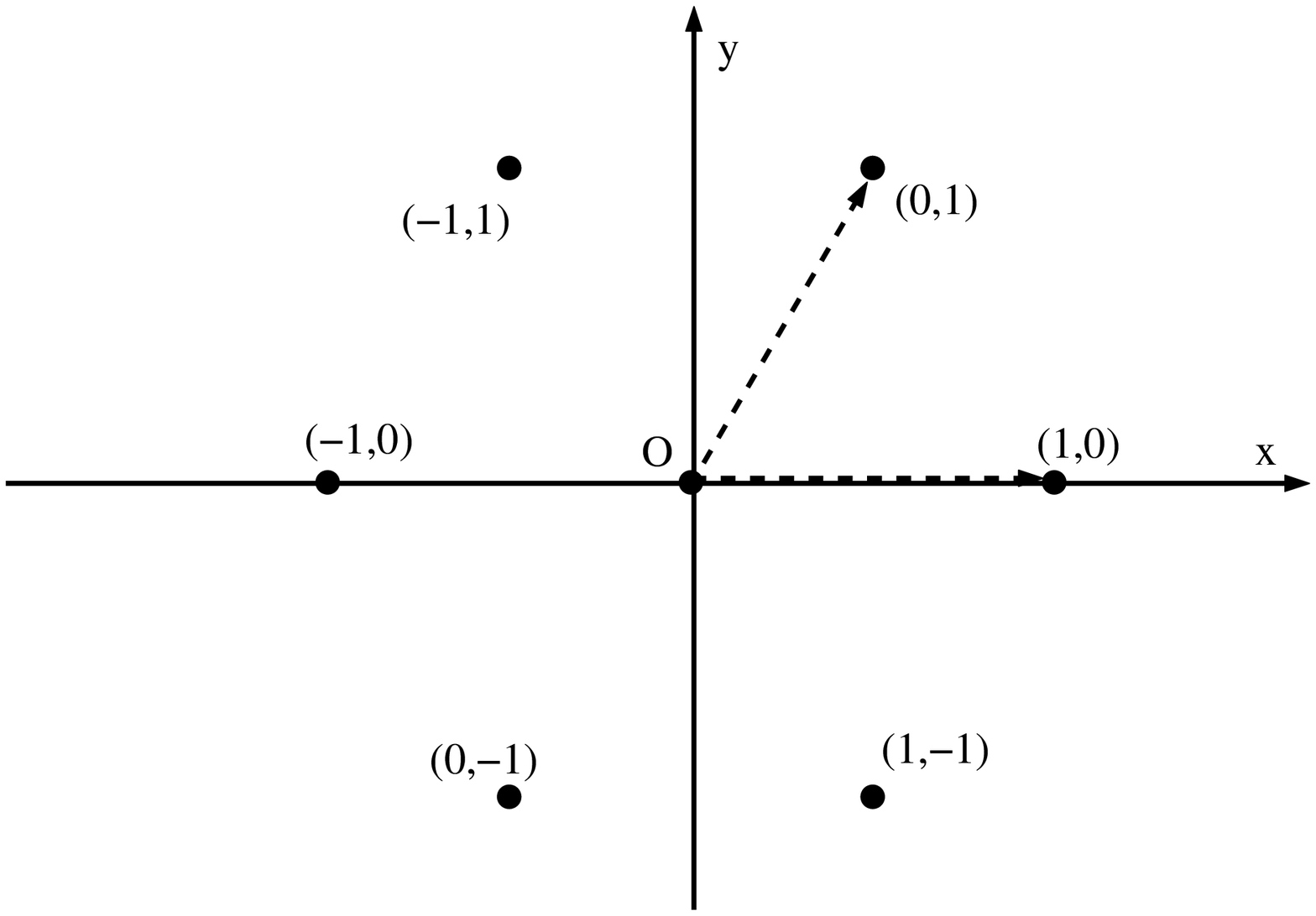}}
\caption{}
\label{tri_lattice}
\end{center}
\end{figure}

\noindent{\bf FIG. \ref{tri_lattice}} %1
Triagular Lattice
\bigskip

\begin{figure} %2
\begin{center}
\resizebox{5.5in}{!}{\includegraphics{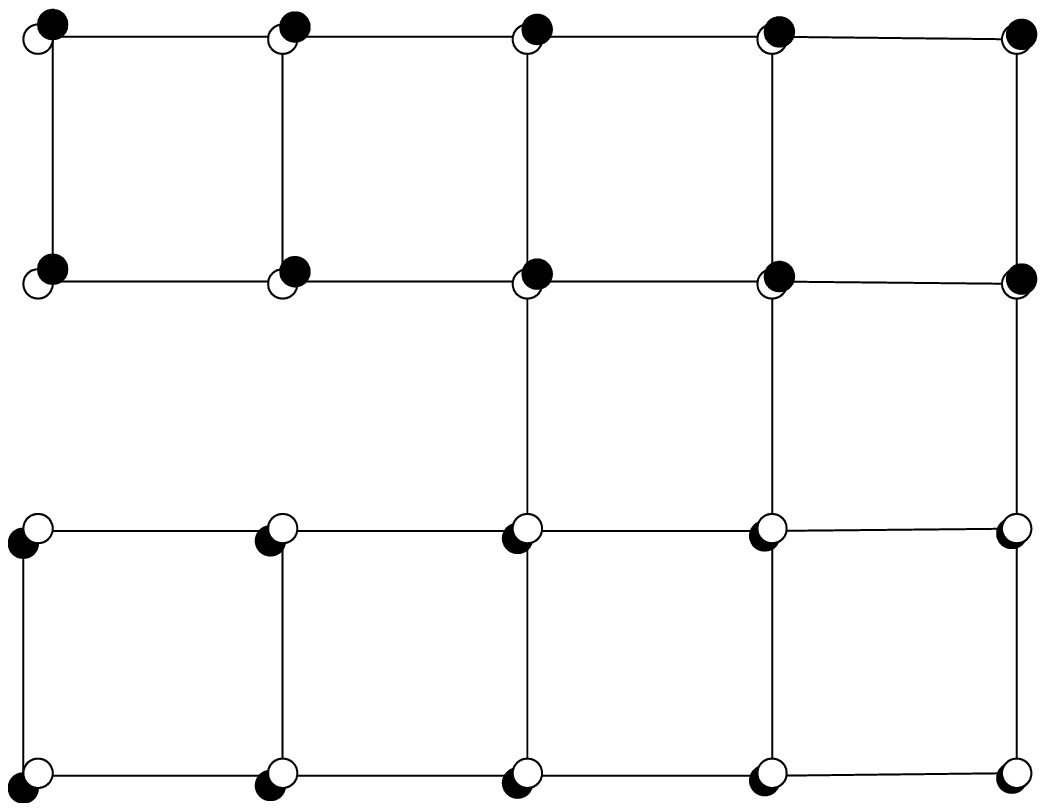}}
\caption{}
\label{sq_lattice}
\end{center}
\end{figure}

\noindent{\bf FIG. \ref{sq_lattice}} %2
2D Slepyan model of fracture. The white dots indicate the equilibrium locations,
the black dots indicate the displaced points once stress is applied.
\bigskip

\begin{figure} %3
\begin{center}
\resizebox{5.5in}{!}{\includegraphics{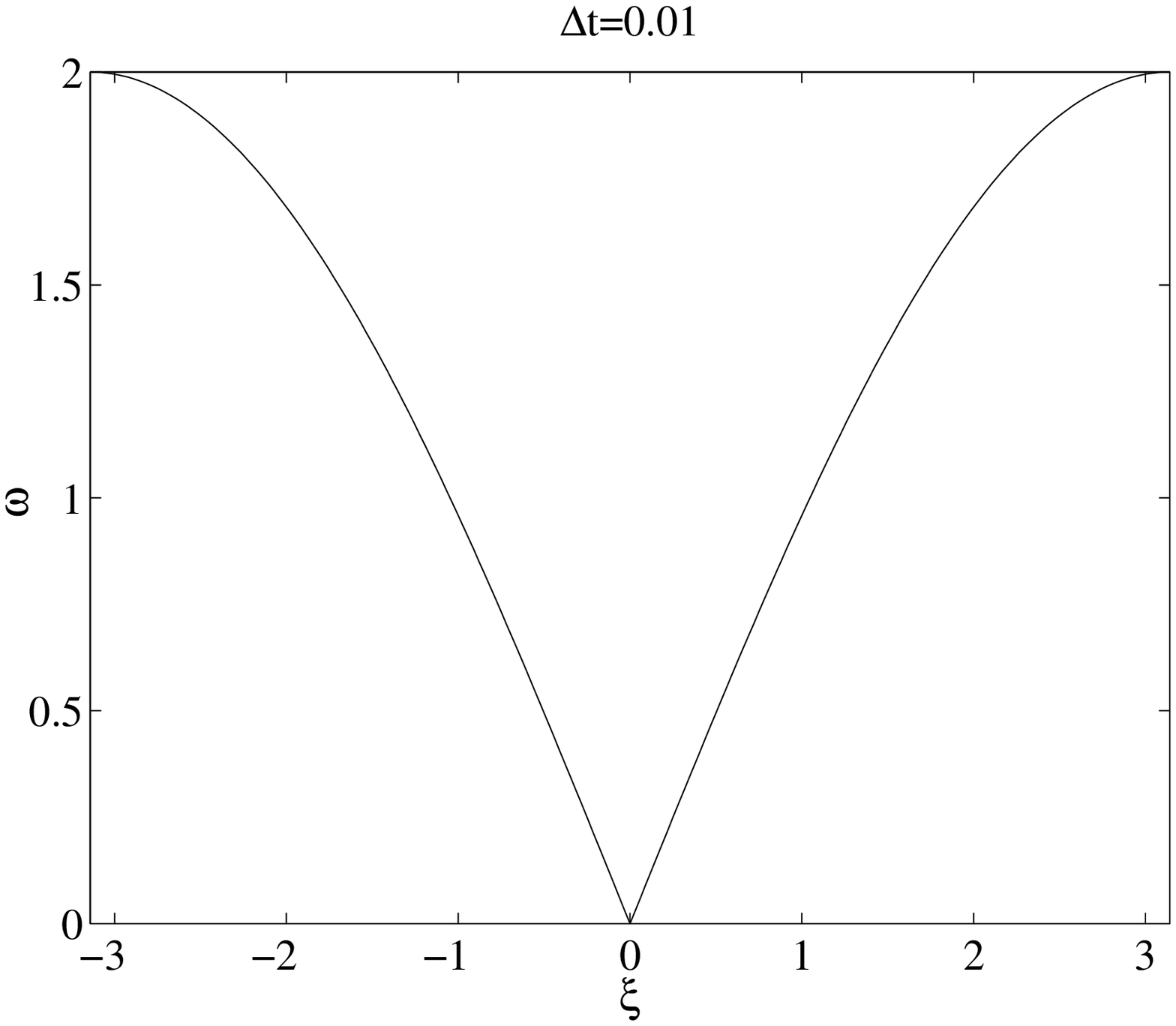}}
\caption{}
\label{1d_dispersion}
\end{center}
\end{figure}

\noindent{\bf FIG. \ref{1d_dispersion}} %3
Dispersion relation
\bigskip

\begin{figure} %4
\begin{center}
\resizebox{5.5in}{!}{\includegraphics{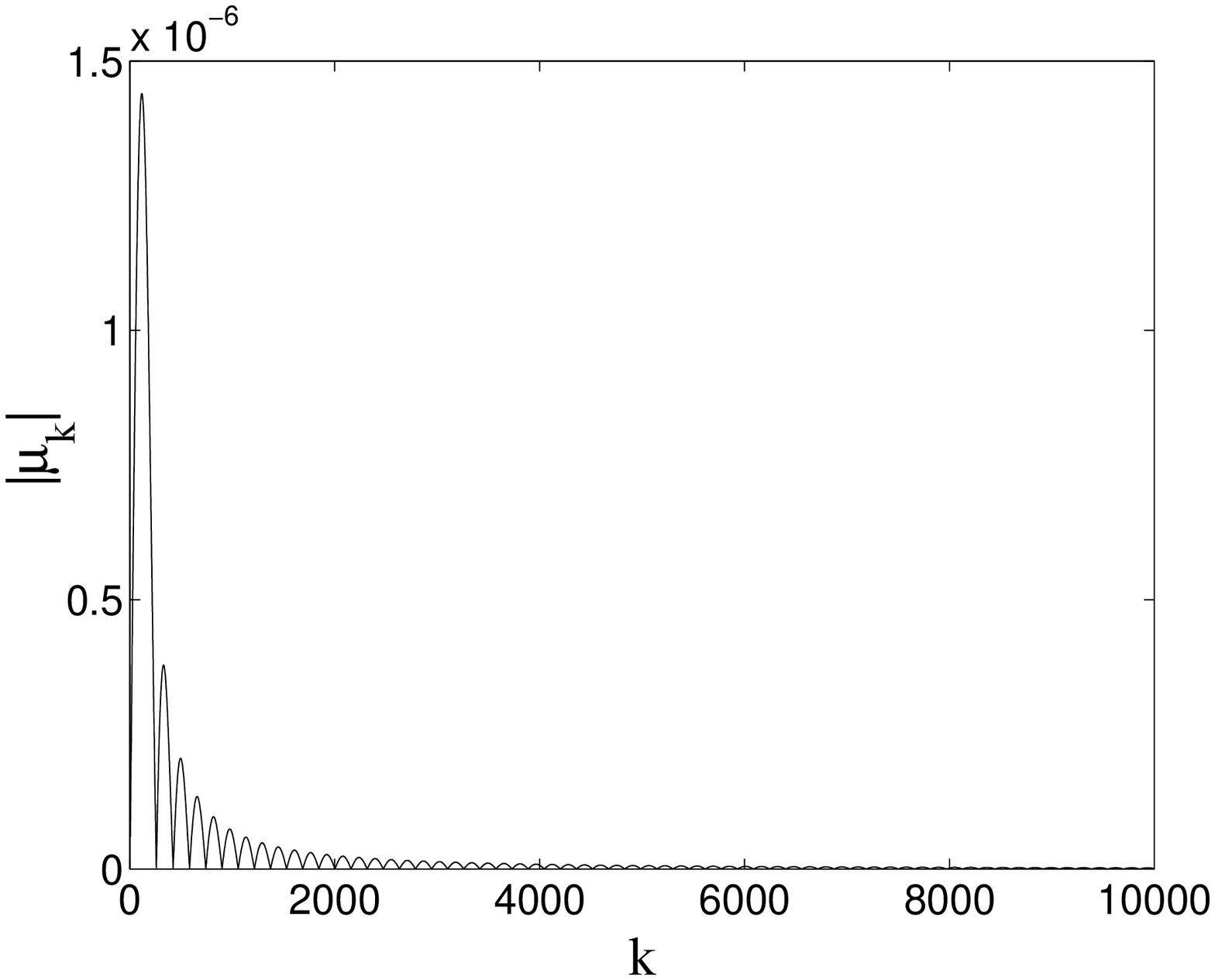}}
\caption{}
\label{fig_mu_decay}
\end{center}
\end{figure}

\noindent{\bf FIG. \ref{fig_mu_decay}} %4
Decay tendency of $|\mu_k|$.
\bigskip

\begin{figure} %5
\begin{center}
\resizebox{5.5in}{!}{\includegraphics{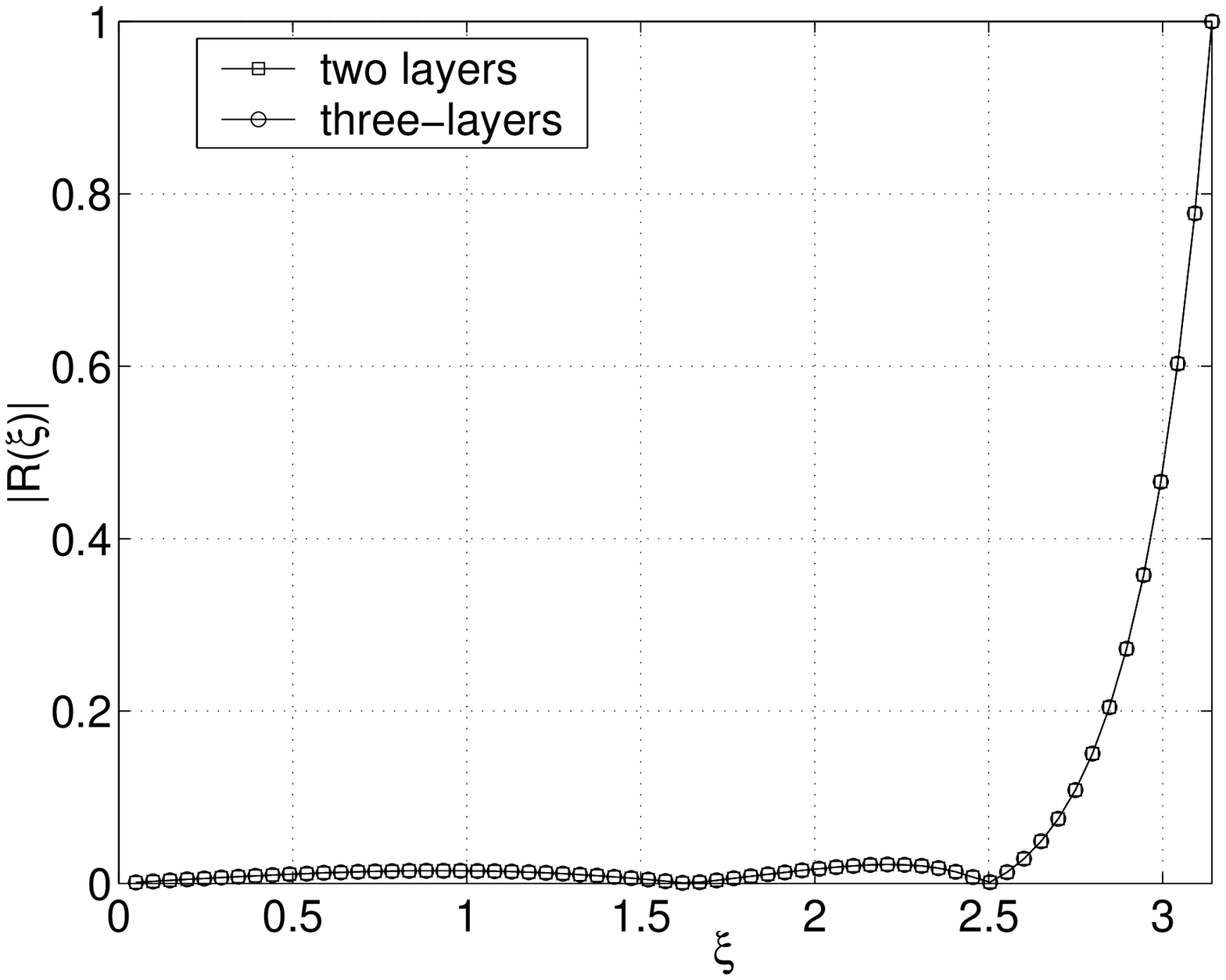}}
\caption{}
\label{reflect_coef}
\end{center}
\end{figure}

\noindent{\bf FIG. \ref{reflect_coef}} %5
Reflection coefficients for (\ref{eq:1d_abc_coef_23}) and 
(\ref{eq:1d_abc_coef_32}).
\bigskip

\begin{figure} %6
\begin{center}
\resizebox{5.5in}{!}{\includegraphics{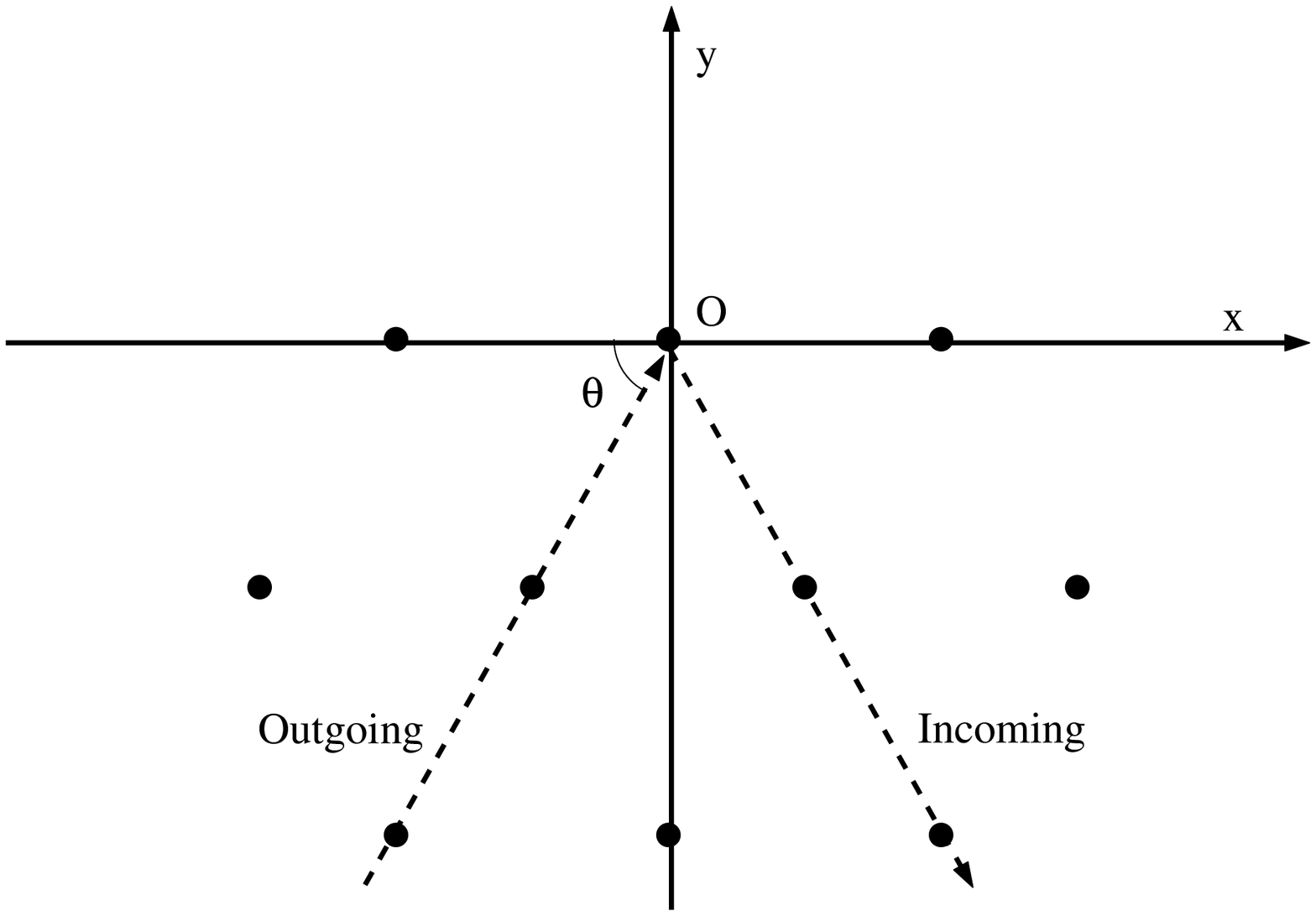}}
\caption{}
\label{fig_reflect}
\end{center}
\end{figure}

\noindent{\bf FIG. \ref{fig_reflect}} %6
The `Incoming' and `Outgoing' phonons near the boundary.
\bigskip

\begin{figure} %7
\begin{center}
\resizebox{5in}{!}{\includegraphics{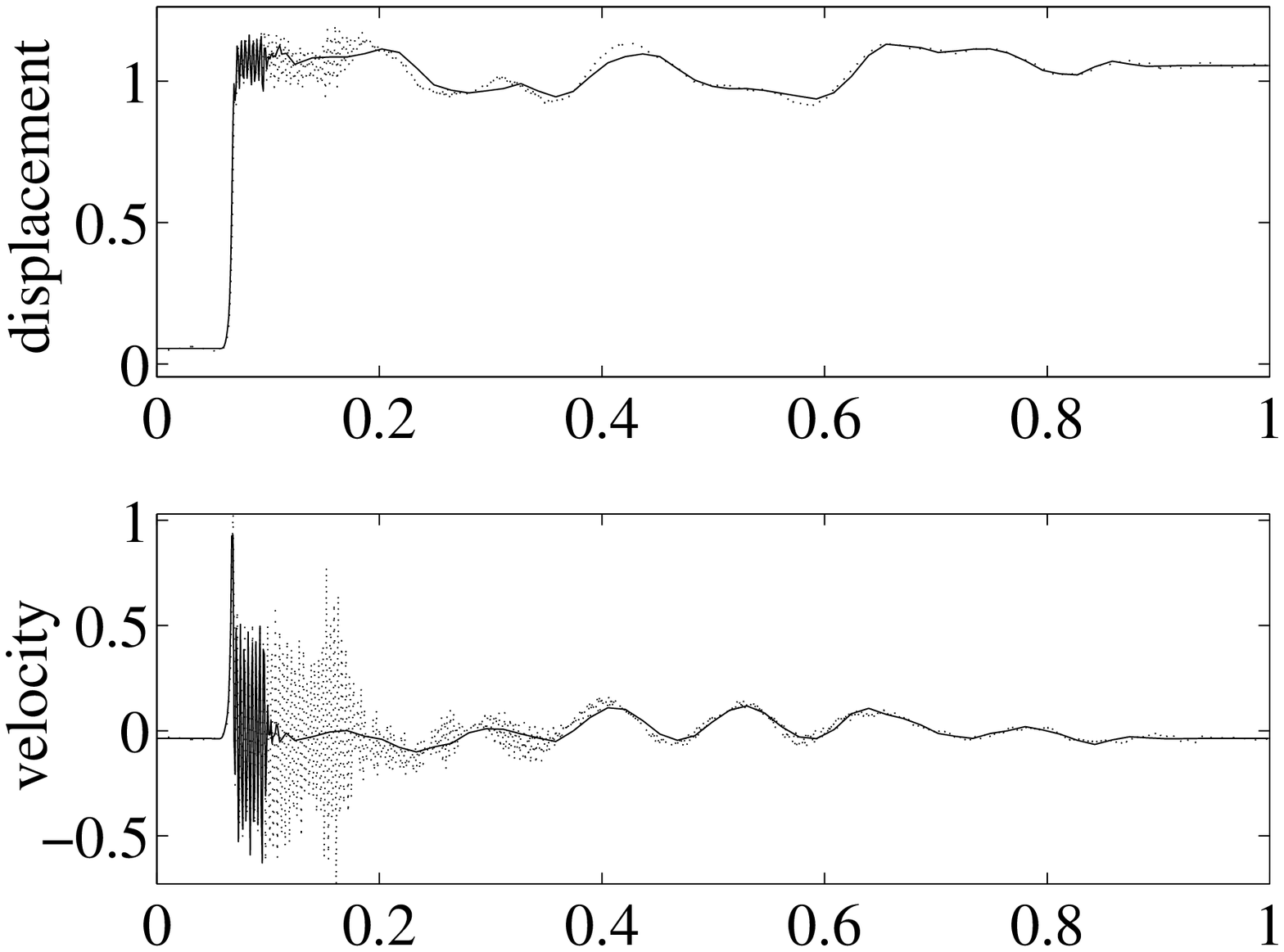}} \vspace{1cm}

\resizebox{5in}{!}{\includegraphics{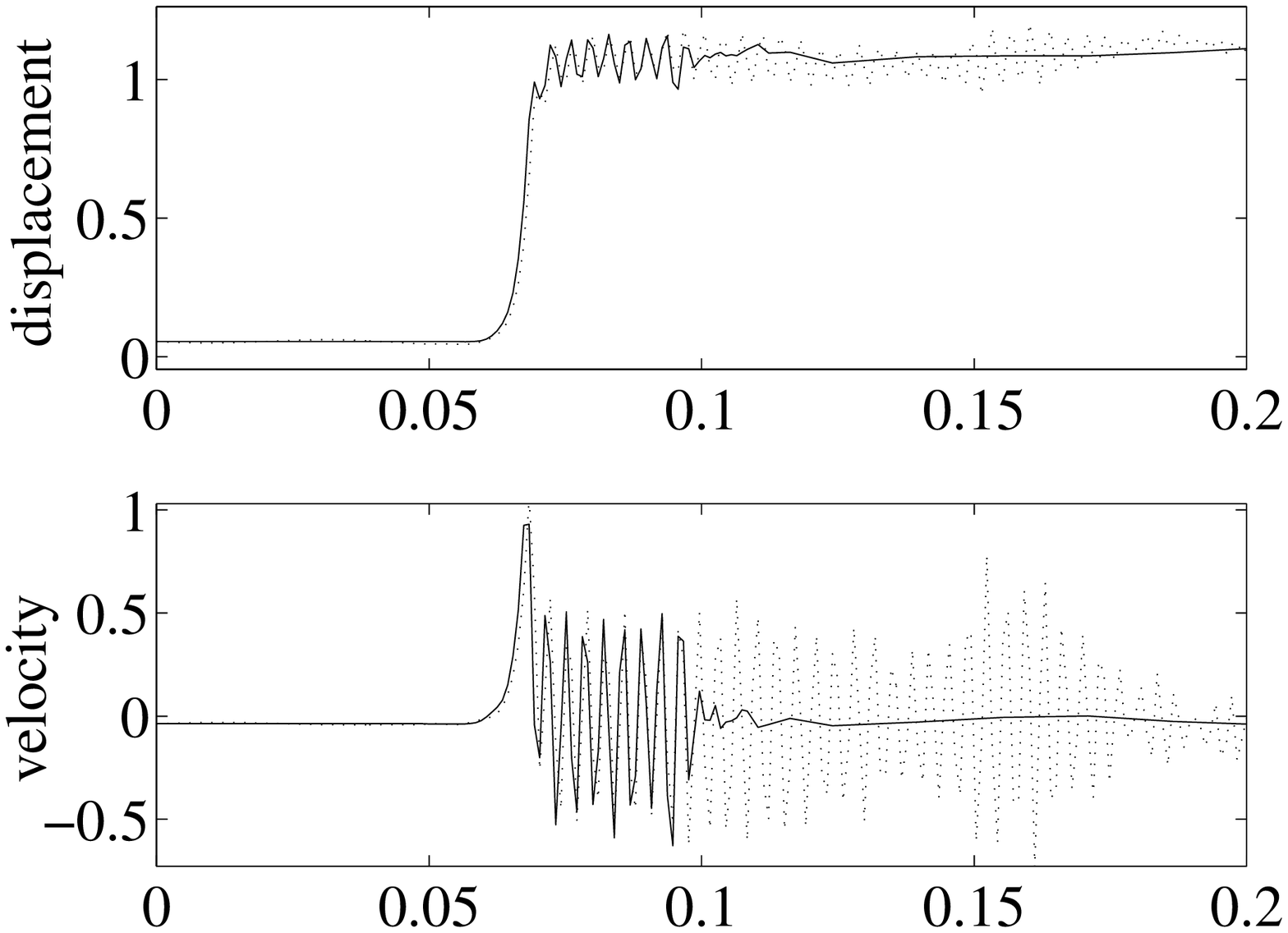}}
\caption{}
\label{fig_ex_fk1}
\end{center}
\end{figure}

\noindent{\bf FIG. \ref{fig_ex_fk1}} %7
Comparison of the displacement and velocity profiles computed
using the full atomistic and the atomistic/continuum models, with $f=0.04$. 
The top two graphs show the results in the whole computational domain. 
The bottom two graphs show the details near the dislocation. The solid line is the
result of the atomistic/continuum method. The dash line is the result
of the full atomistic method.
\bigskip

\begin{figure} %8
\begin{center}
\resizebox{5in}{!}{\includegraphics{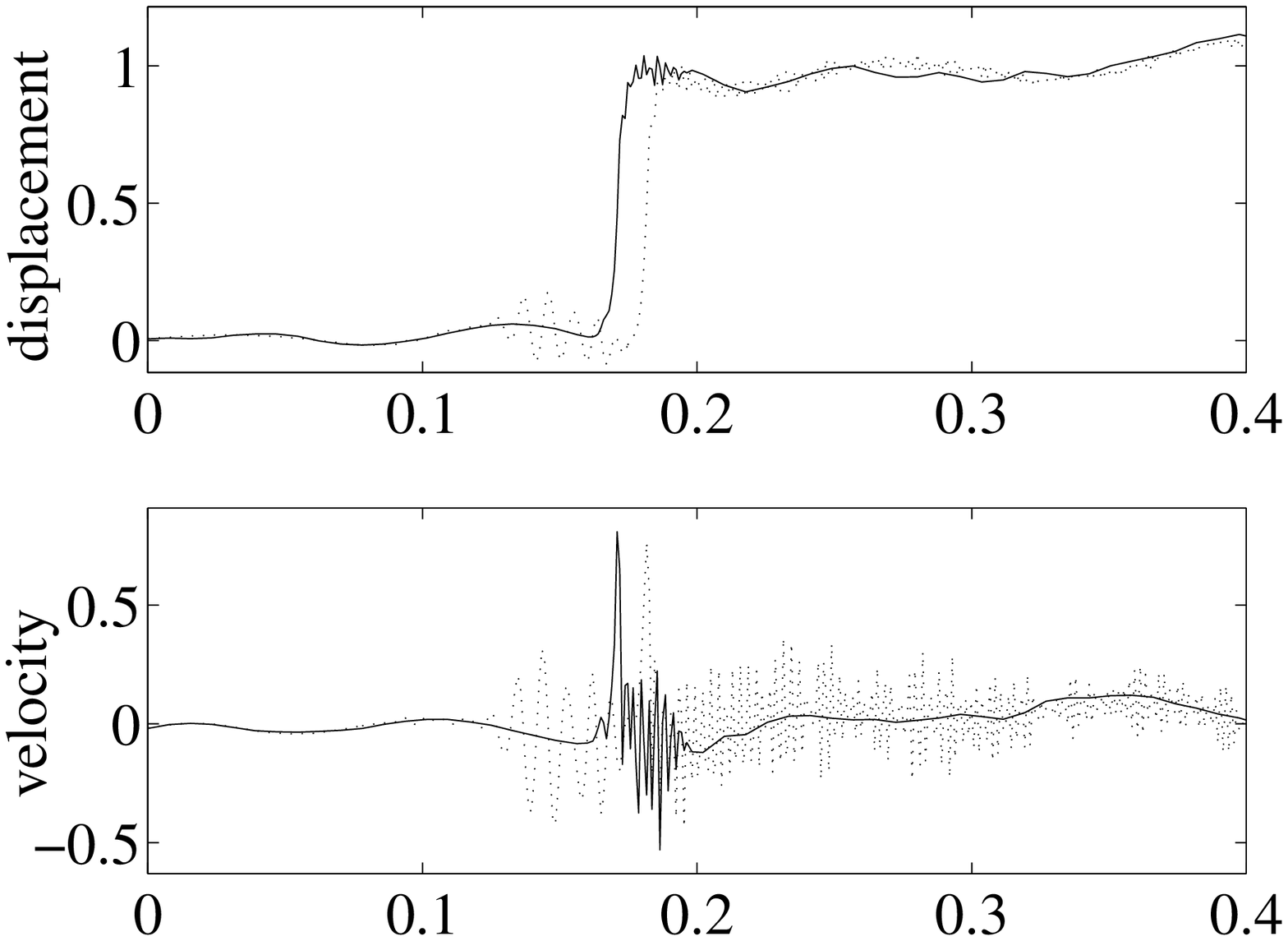}} \vspace{1cm}

\resizebox{5in}{!}{\includegraphics{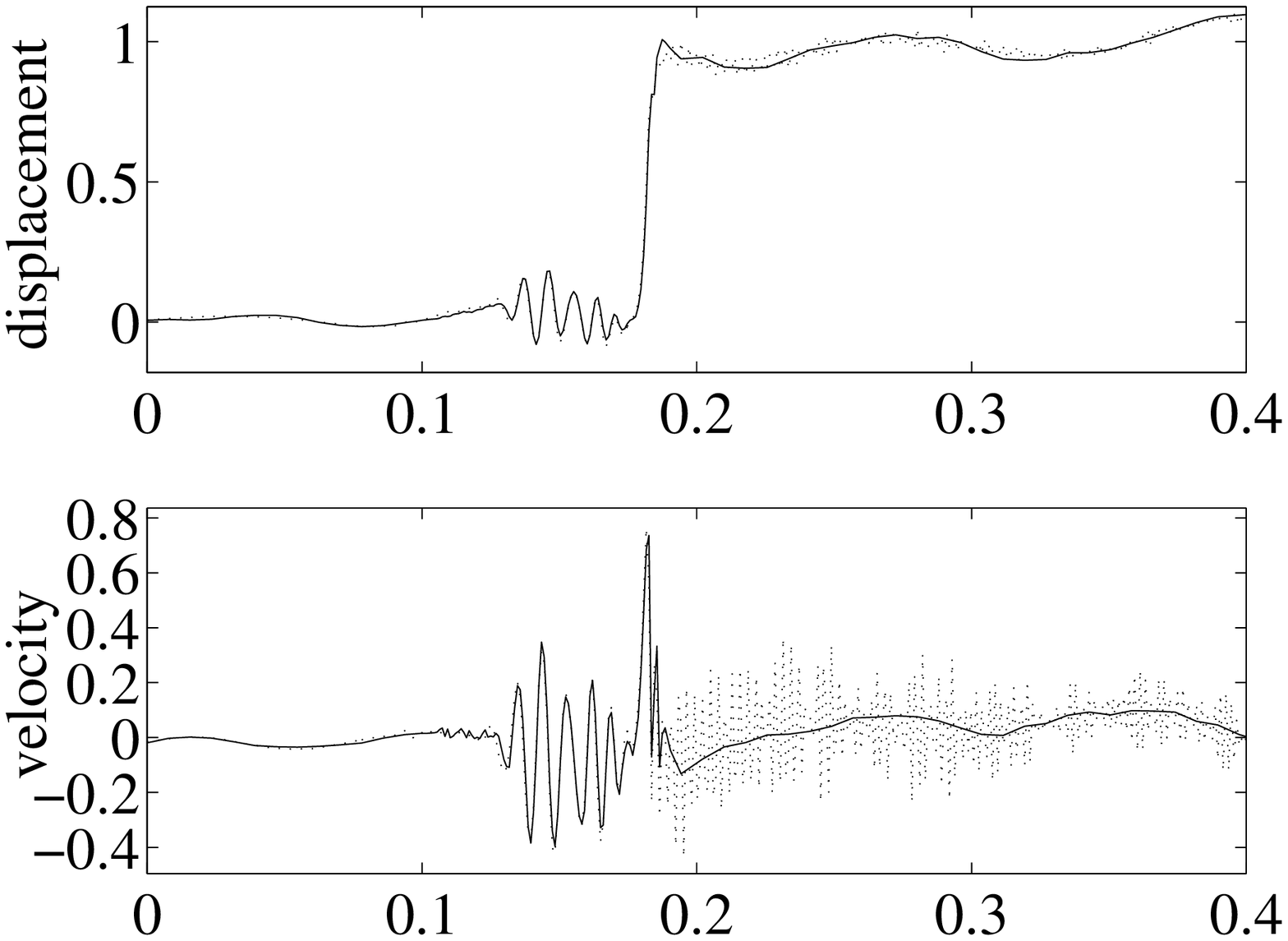}}
\caption{}
\label{fig_ex_fk2}
\end{center}
\end{figure}

\noindent{\bf FIG. \ref{fig_ex_fk2}} %8
Comparison of the displacement and velocity profiles computed
using the full atomistic and the atomistic/continuum models, with $f=0.02$.
The top two graphs show the results when the transition
from the atomistic to continuum regions is sharp. The bottom two graphs
show the results when the transition is gradual. Solid line is the
result of the atomistic/continuum method. The dash line is the result
of the full atomistic method. Only the region near the dislocation is shown.
\bigskip

\begin{figure} %9
\begin{center}
\resizebox{4.5in}{!}{\includegraphics{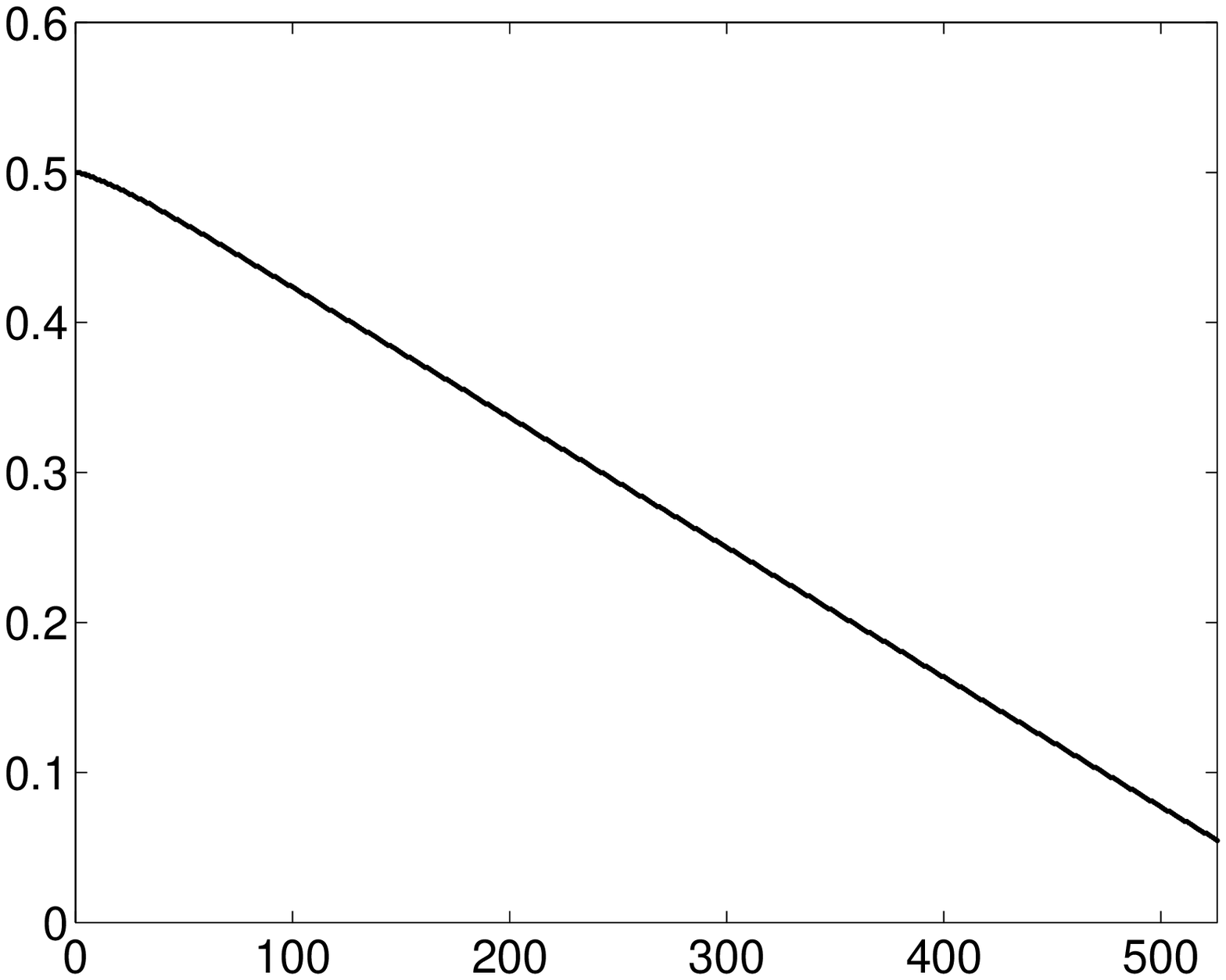}} \vspace{1cm}

\resizebox{4.5in}{!}{\includegraphics{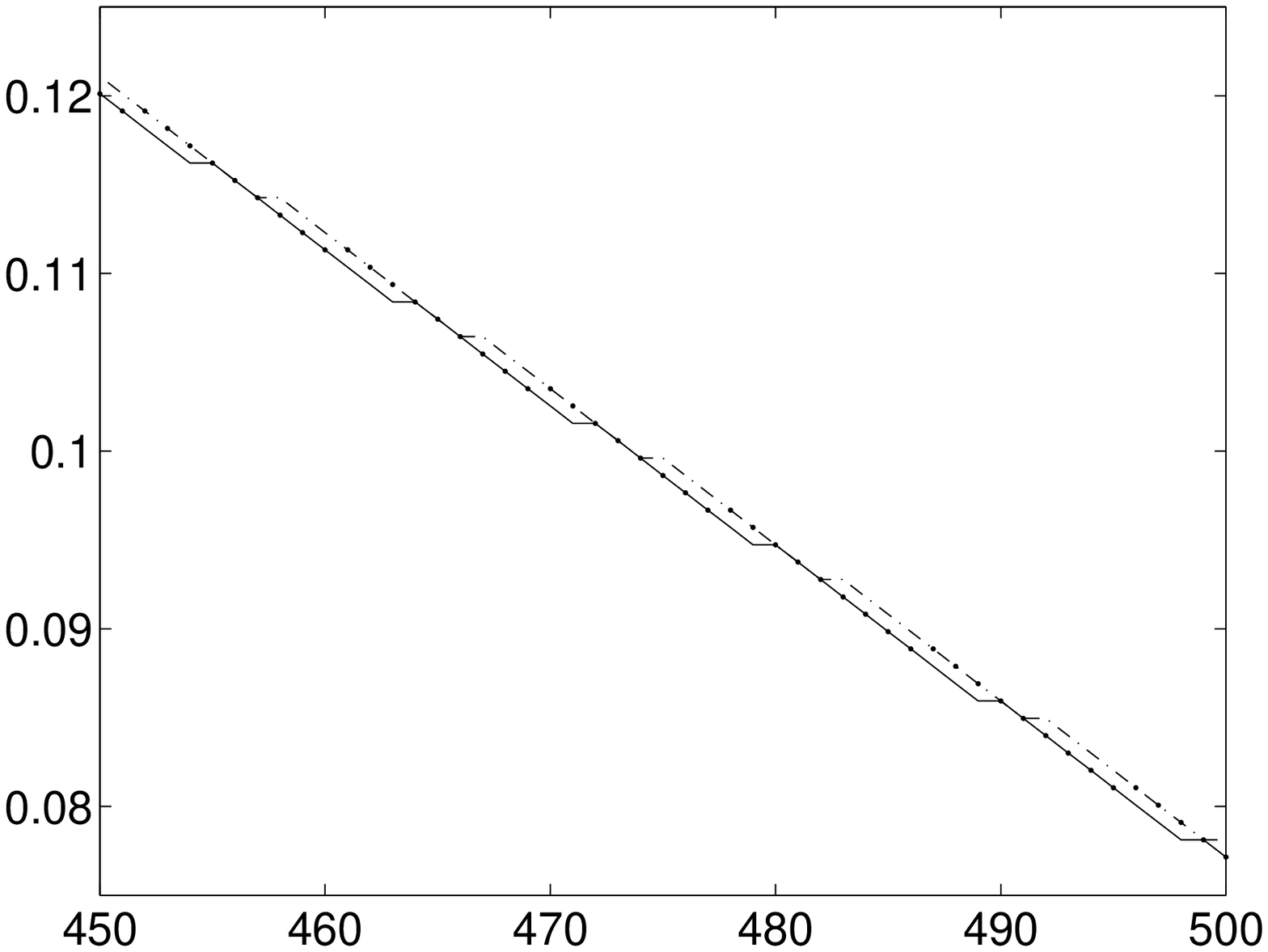}}
\caption{}
\label{dislocation_pos_04}
\end{center}
\end{figure}

\noindent{\bf FIG. \ref{dislocation_pos_04}} %9
Comparison of the positions of the dislocation as a function of time computed
using the coupled method and the detailed molecular dynamics with $f=0.04$.
Dot line is the result of full MD simulation; solid line is the result with
gradual transition between atomistic and continuum regions; dash line is the 
result with sharp transition.
\bigskip

\begin{figure} %10
\begin{center}
\resizebox{5in}{!}{\includegraphics{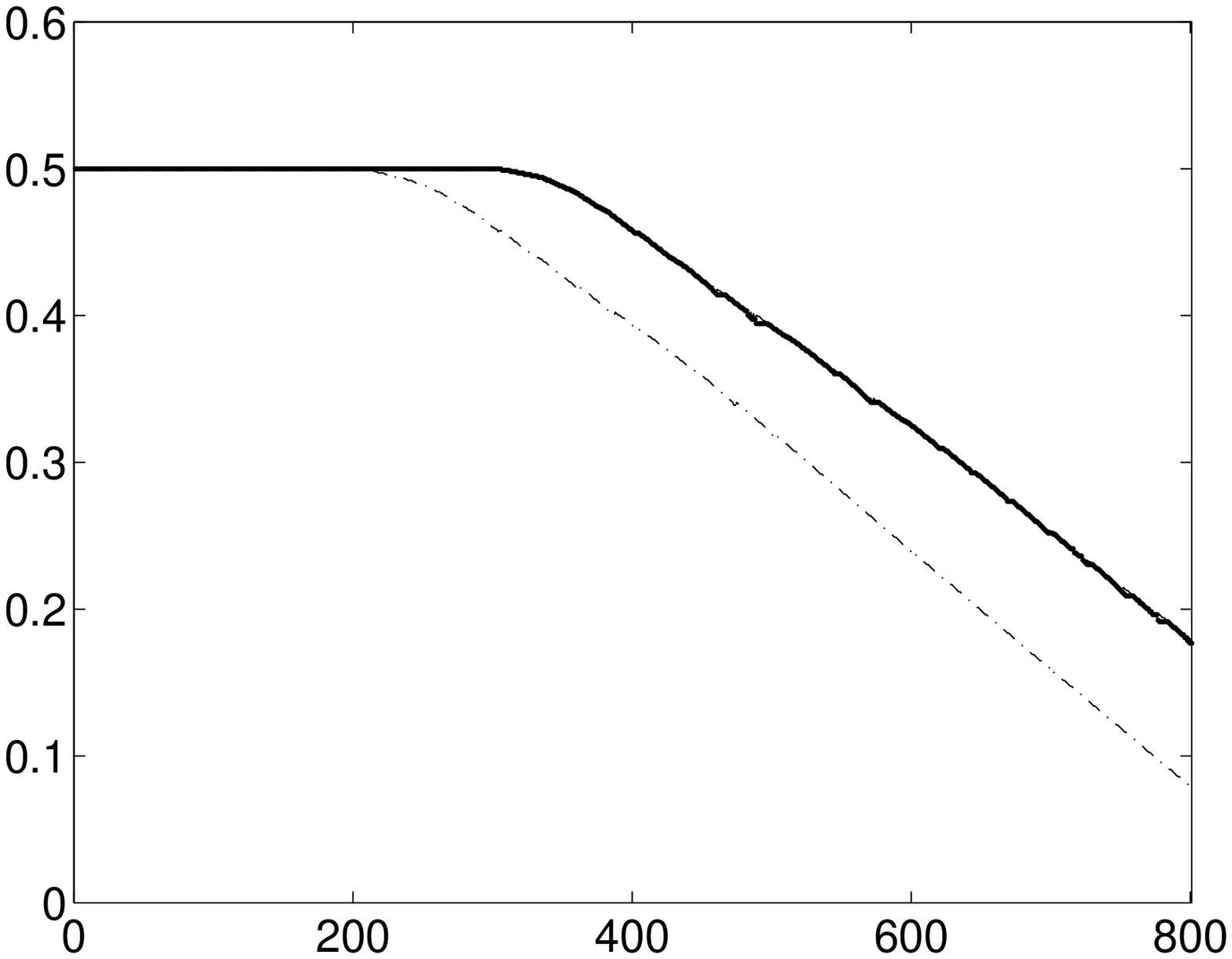}} \vspace{1cm}

\resizebox{5in}{!}{\includegraphics{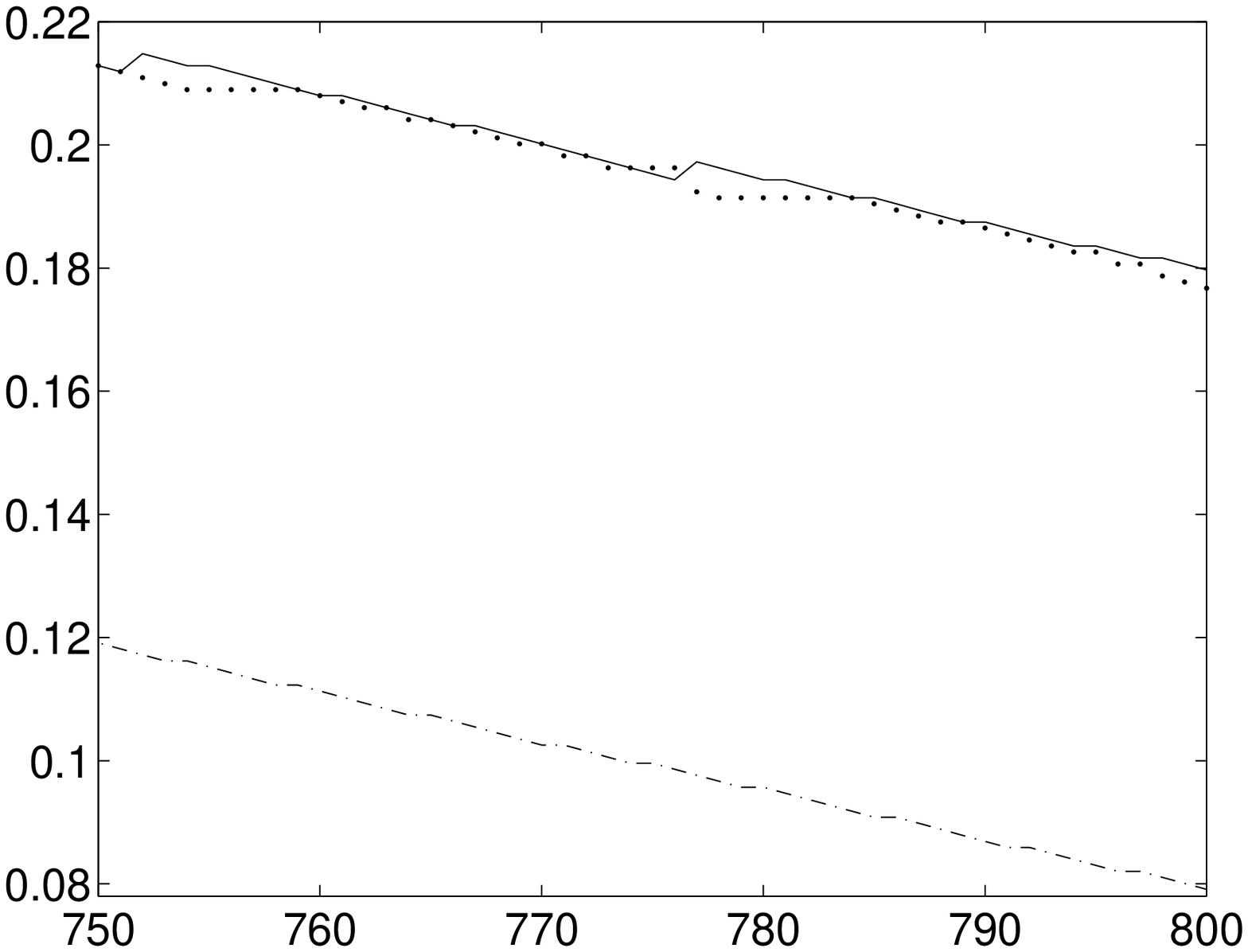}}
\caption{}
\label{dislocation_pos_02}
\end{center}
\end{figure}

\noindent{\bf FIG. \ref{dislocation_pos_02}} %10
Comparison of the positions of the dislocation as a function of time computed
using the coupled method and the detailed molecular dynamics with $f=0.02$.
Dot line is the result of full MD simulation; solid line is the result with
gradual transition between atomistic and continuum regions; dash line is the 
result with sharp transition.
\bigskip

\begin{figure} %11
\begin{center}
\resizebox{5.5in}{!}{\includegraphics{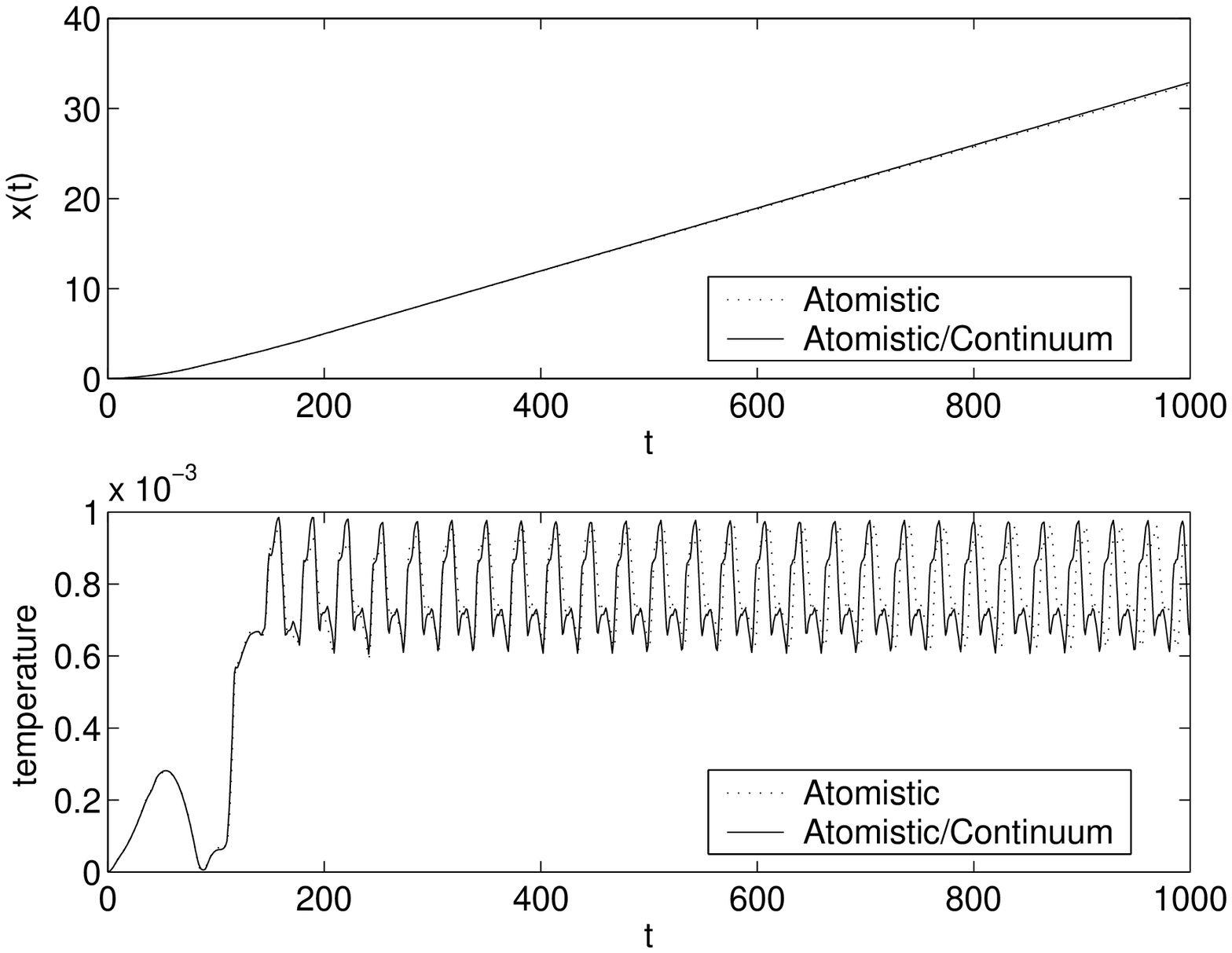}}
\caption{}
\label{fig_flat_fric}
\end{center}
\end{figure}

\noindent{\bf FIG. \ref{fig_flat_fric}} %11
Displacement and temperature as a function of time for the friction
problem. 
\bigskip

\begin{figure} %12
\begin{center}
\resizebox{5.5in}{!}{\includegraphics{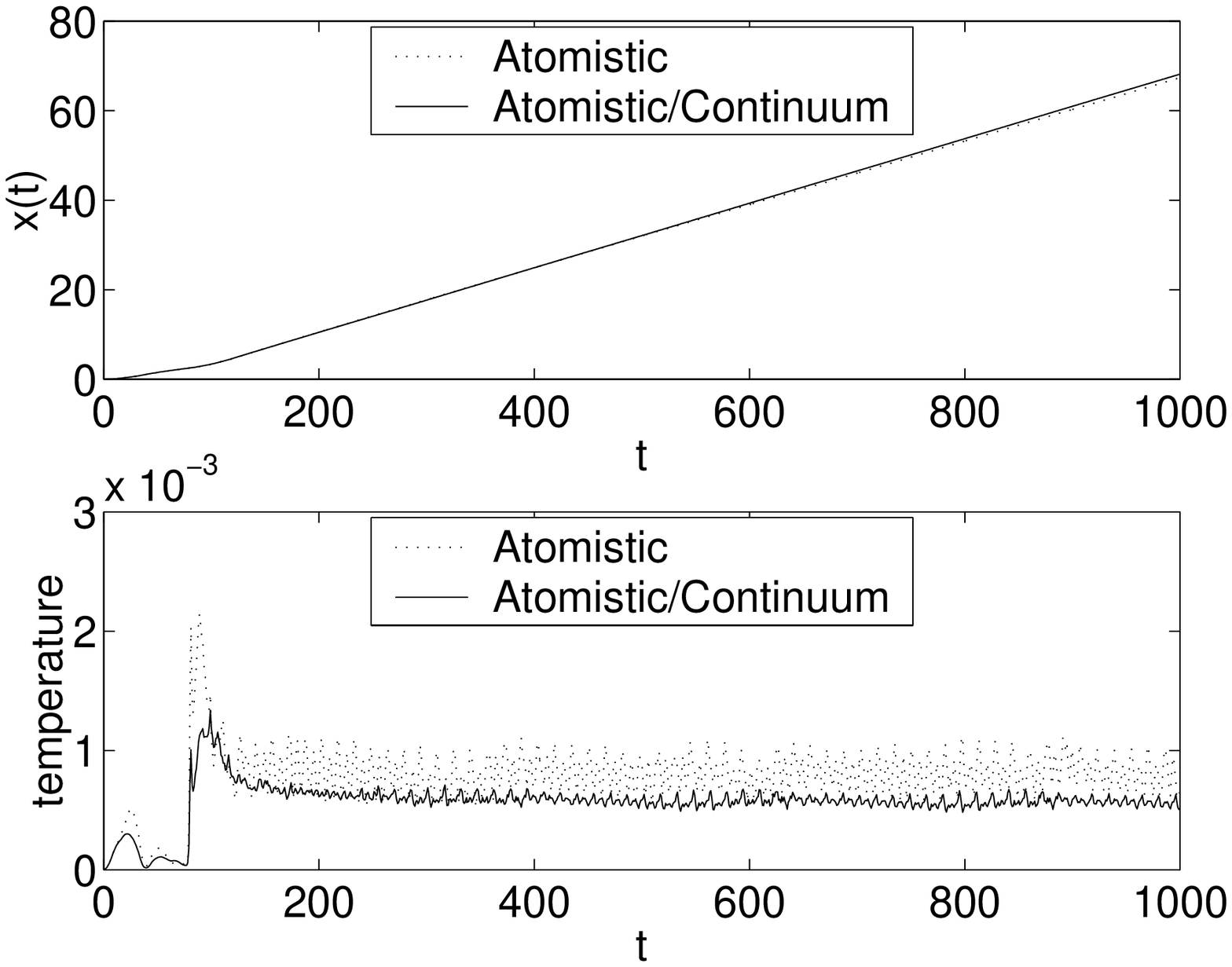}}
\caption{}
\label{fig_rough_fric}
\end{center}
\end{figure}

\noindent{\bf FIG. \ref{fig_rough_fric}} %12
Displacement and temperature as a function of time for friction between 
rough surfaces.
\bigskip

\begin{figure} %13
\begin{center}
\resizebox{5in}{!}{\includegraphics{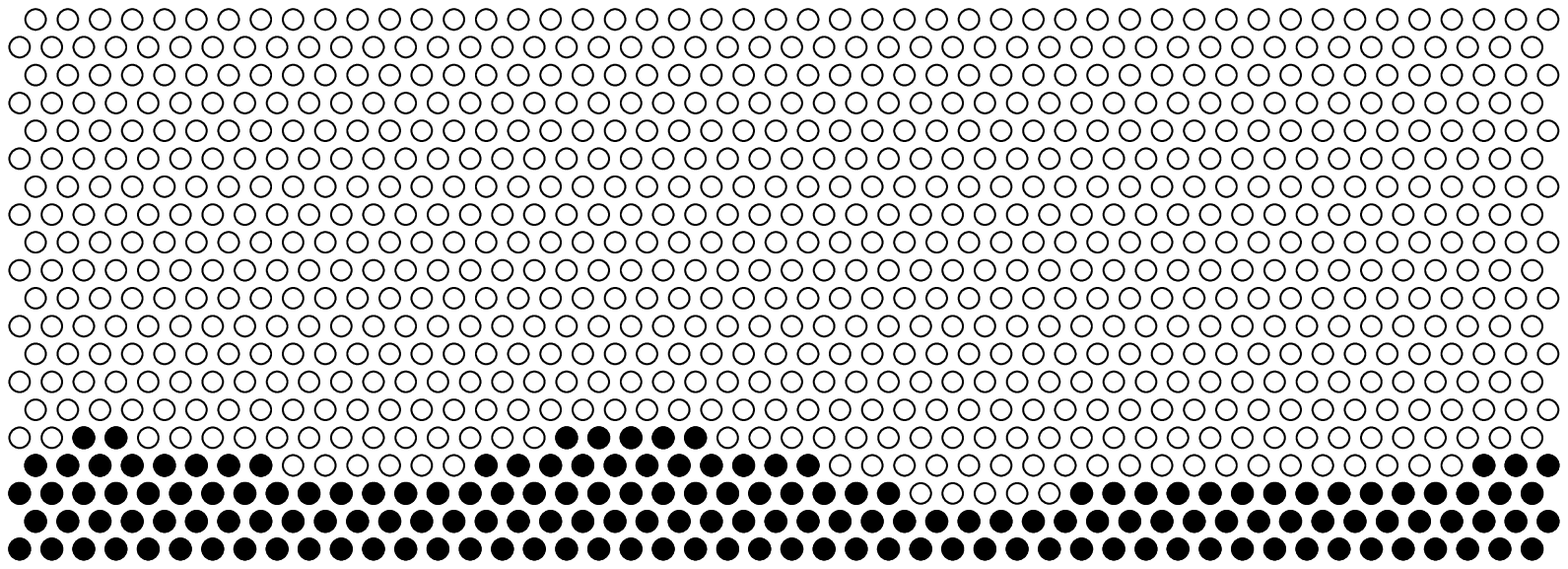}} \vspace{1cm}

\resizebox{5in}{!}{\includegraphics{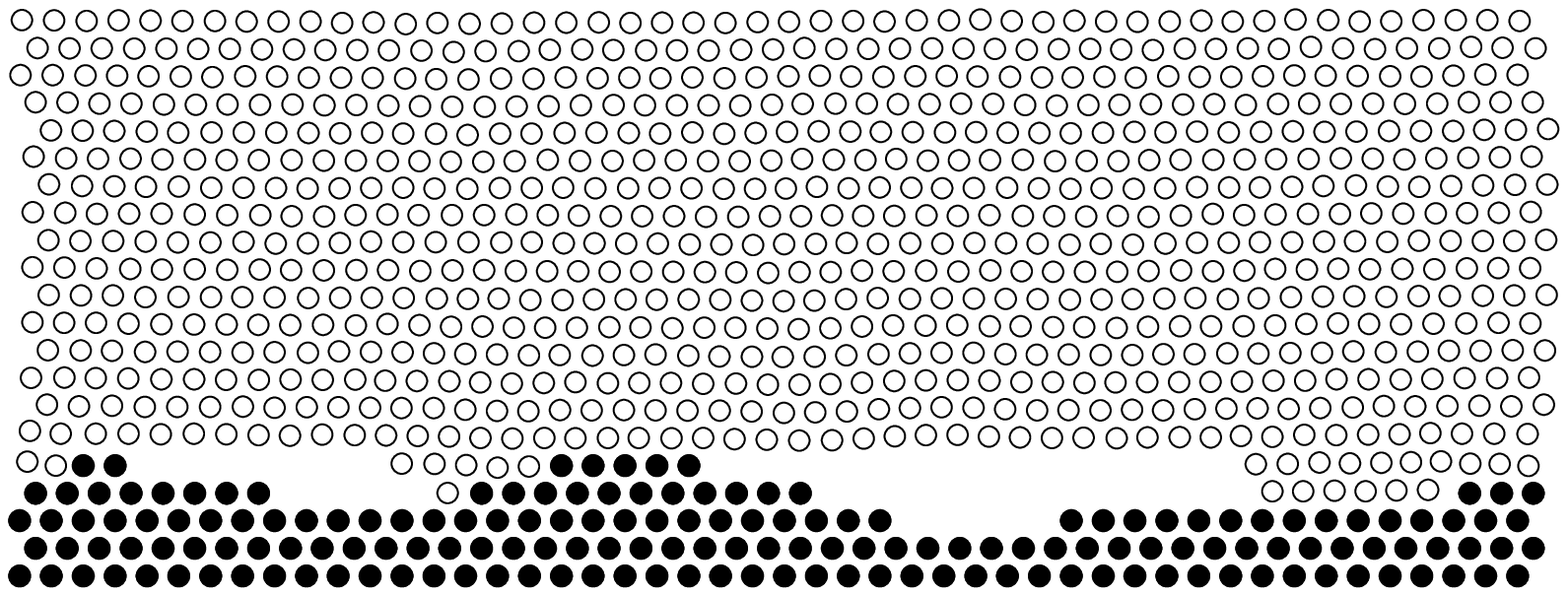}}
\caption{}
\label{friction_interface}
\end{center}
\end{figure}

\noindent{\bf FIG. \ref{friction_interface}} %13
The positions of the atoms near the interfaces. The white circles are light
atoms, the black ones are heavy atoms. The top graph is the initial state,
the bottom graph is the late state at $t=1000$.
\bigskip

\begin{figure} %14
\begin{center}
\resizebox{5in}{!}{\includegraphics{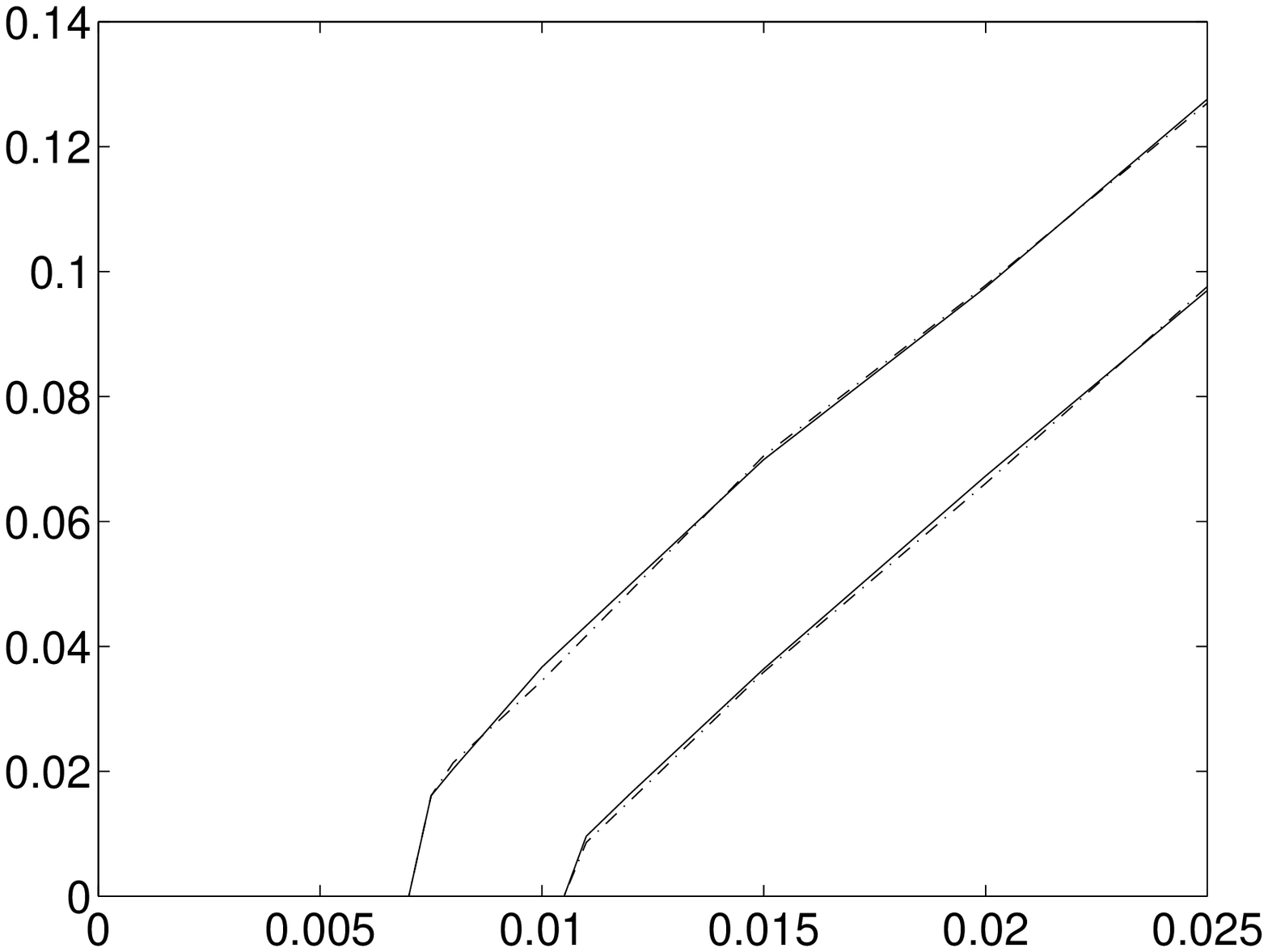}}

\caption{}
\label{vel_force}
\end{center}
\end{figure}

\noindent{\bf FIG. \ref{vel_force}} %14
Comparison of the force-velocity relations for both flat and rough interfaces.
The top two lines are the results for flat case,
the bottom two lines are the results for rough case.
The solid lines are the results for coupled atomistic/continuum method, 
the dash lines are the results for full MD simulation.
\bigskip

\begin{figure} %15
\begin{center}
\resizebox{5.5in}{!}{\includegraphics{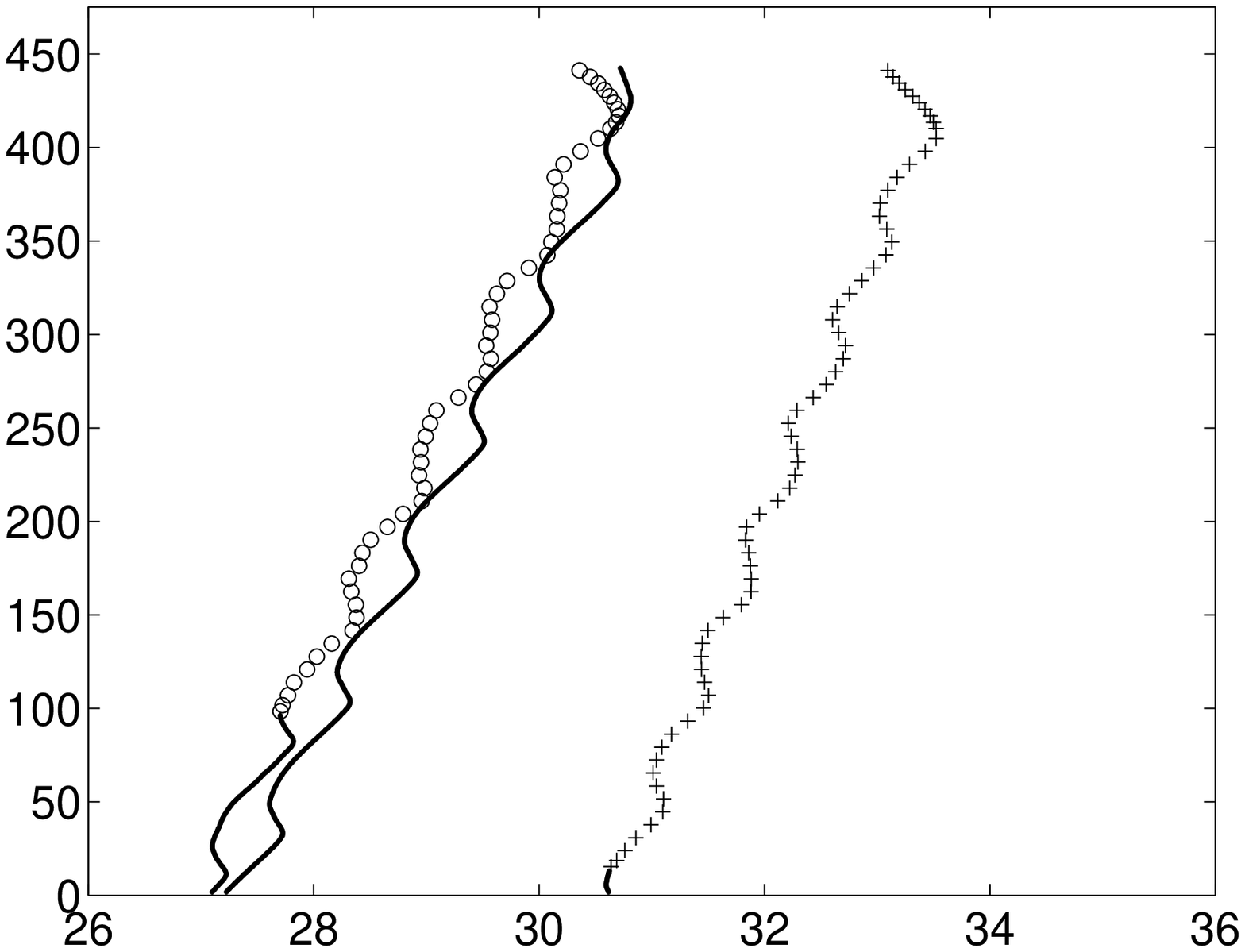}}
\caption{}
\label{fig_friction_1d}
\end{center}
\end{figure}

\noindent{\bf FIG. \ref{fig_friction_1d}} %15
Comparison of the atomic positions of a column of atoms which had the same 
$x$-coordinates initially. Solid line is the result of full MD, the line with
$\circ$ is the result of coupled method with 96 layers in the atomistic region,
the line with $+$ is the result of coupled method with 16 layers in the atomistic region.
In the coupled method, the ratio of atomistic and continuum grids is 1:8.
\bigskip

\begin{figure} %16
\begin{center}
\resizebox{7.5cm}{!}{\includegraphics{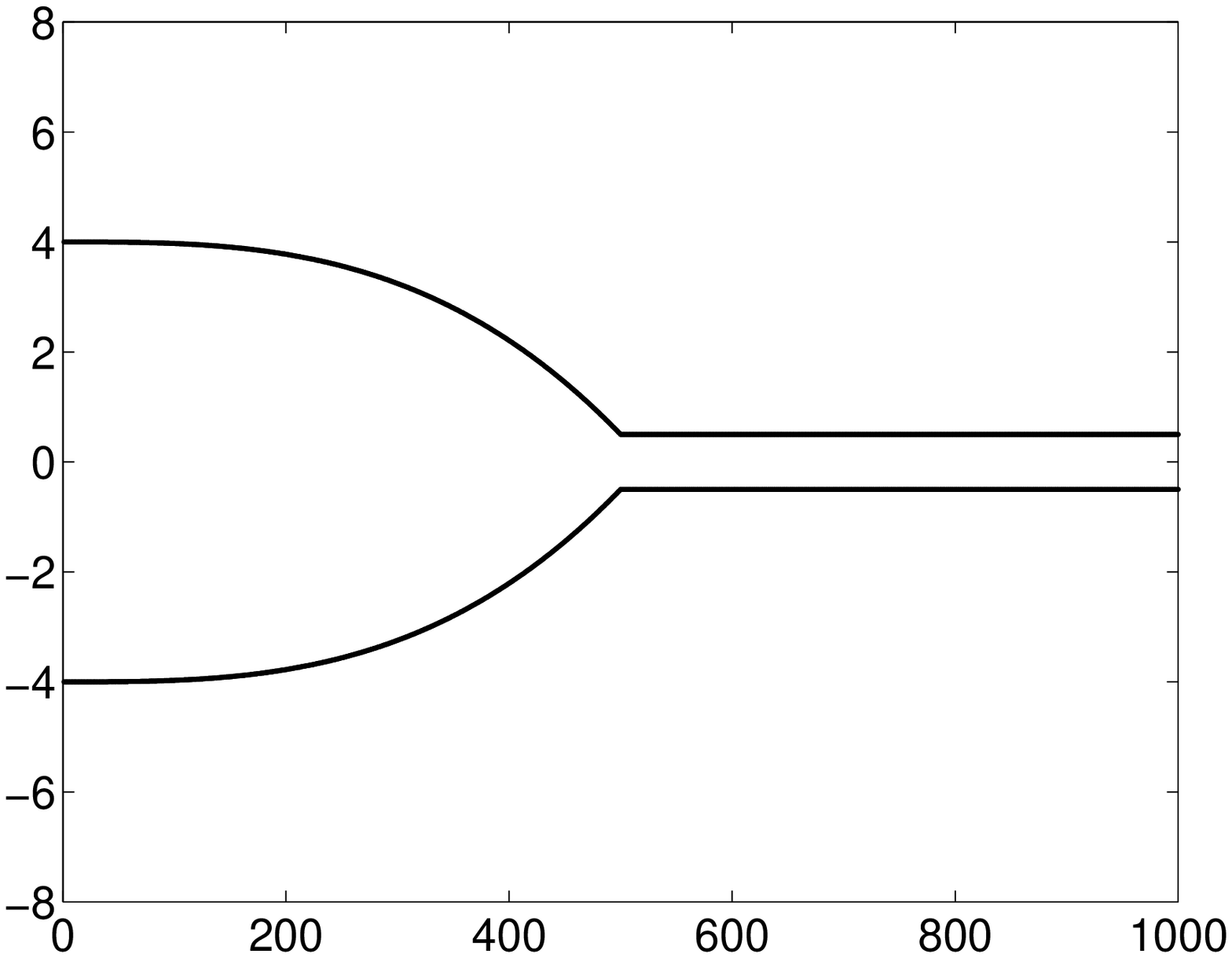}}
\resizebox{7.5cm}{!}{\includegraphics{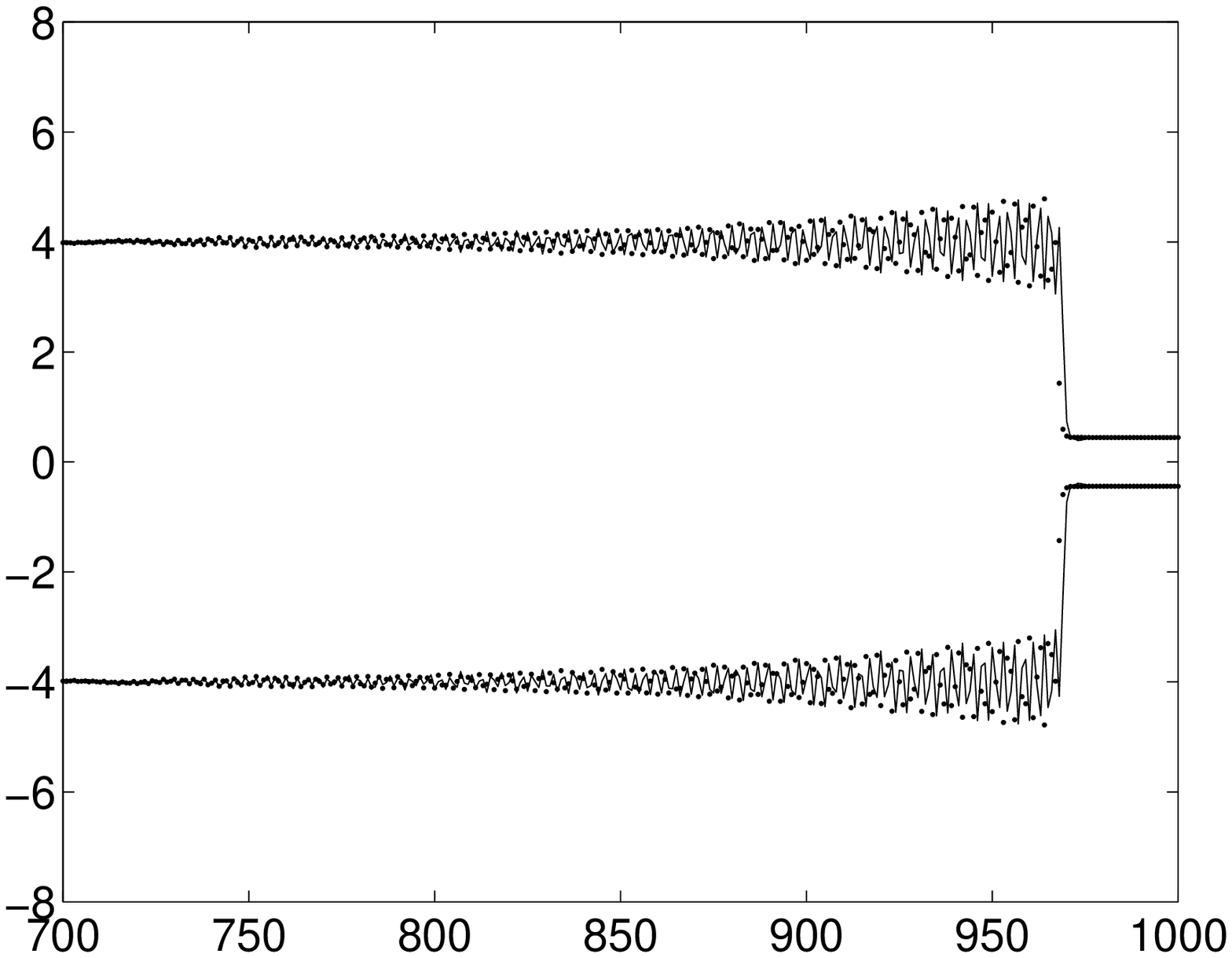}}
\caption{}
\label{fig_1d_fracture}
\end{center}
\end{figure}

\noindent{\bf FIG. \ref{fig_1d_fracture}} %16
One-dimensional fracture problem.
The left graph shows the fracture surface at time $t=0$ and the 
right one shows the fracture surface at time $t=600$ with $b=0.01$, $N=U_N=4$.
The dash line is the result of the full molecular dynamics simulation. The solid 
line is the result of the coupled atomistic/continuum method.
The ratio of atomistic and continuum grids is 1:8.
\bigskip

\begin{figure} %17
\begin{center}
\resizebox{7.5cm}{!}{\includegraphics{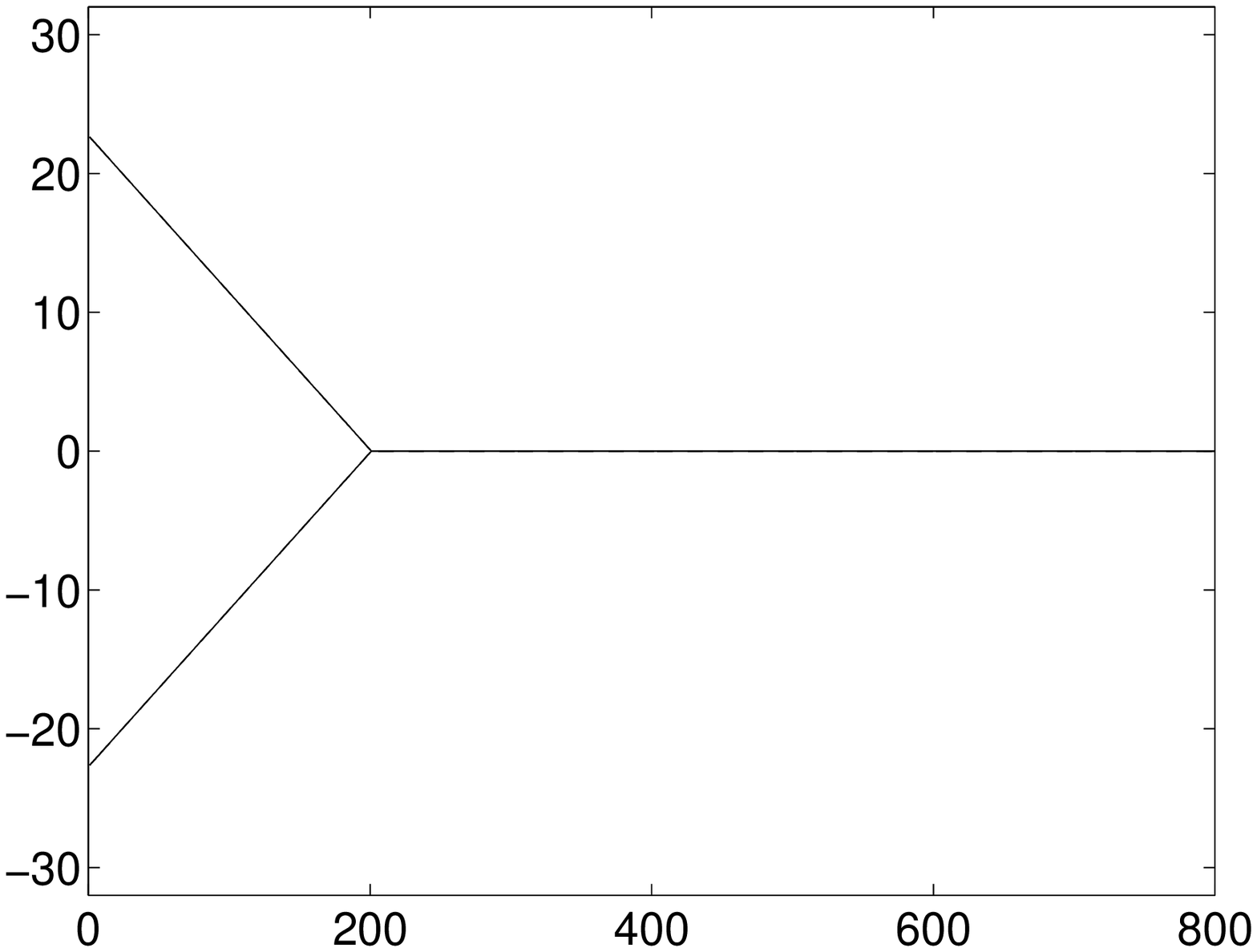}}
\resizebox{7.5cm}{!}{\includegraphics{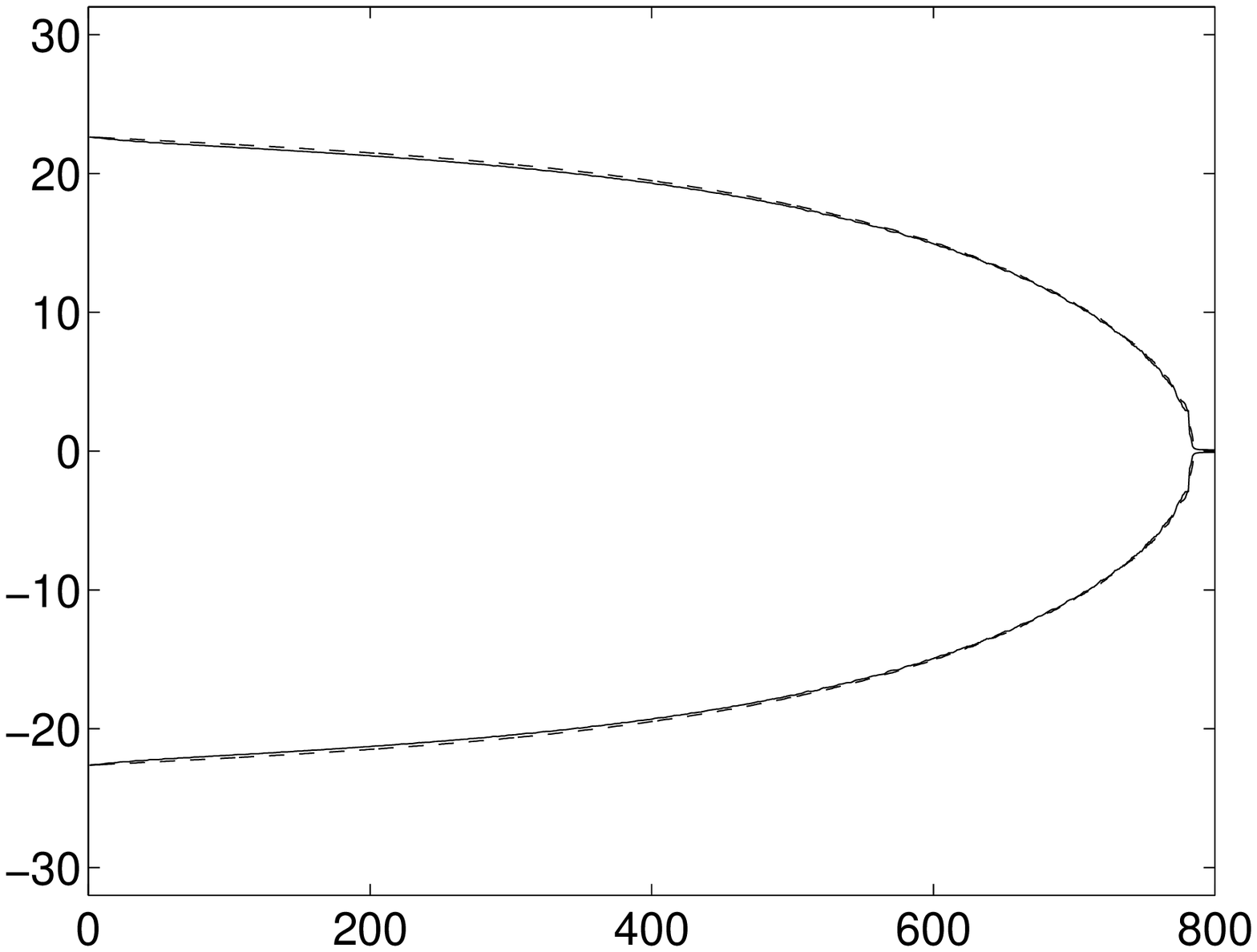}}
\caption{}
\label{fig_2d_fracture}
\end{center}
\end{figure}

\noindent{\bf FIG. \ref{fig_2d_fracture}} %17
Two-dimensional fracture problem.
The left graph shows the fracture surface at time $t=0$ and the 
right one shows the fracture surface at time $t=200$ with $b=0.01$, $N=512$,
and $U_N=\sqrt{N}$. There are 800 atoms in each row.
The dash line is the result of the full molecular dynamics simulation. The solid 
line is the result of the coupled atomistic/continuum method.
In coupled method,
we divide the whole domain into three parts. The middle part including the
crack surface with $800 \tm 64$ atoms is the MD region. The top and bottom parts
are continuum regions.
The ratio of atomistic and continuum grids in each dimension is 1:8.
\bigskip

\begin{figure} %18
\begin{center}
\resizebox{5in}{!}{\includegraphics{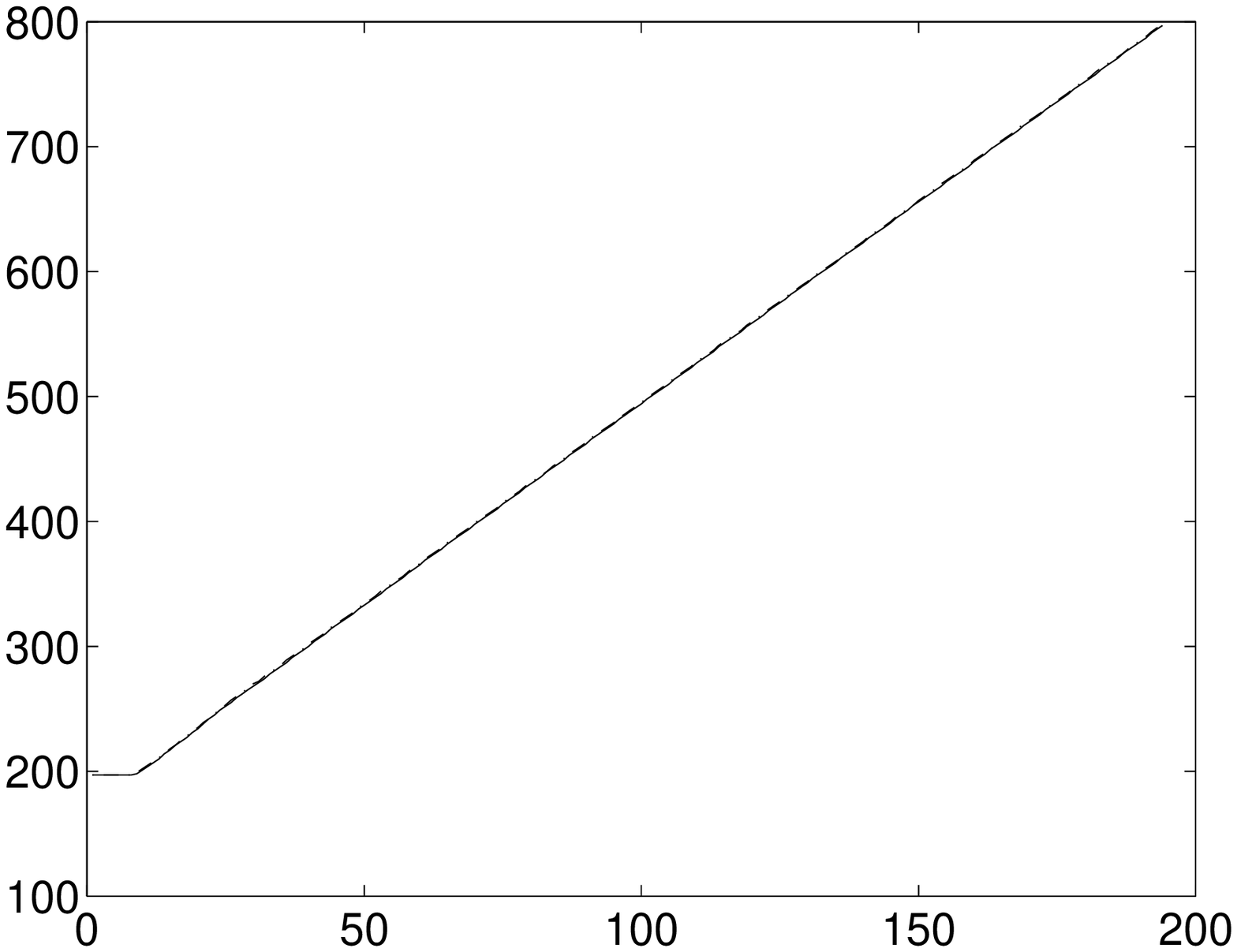}}
\caption{}
\label{crack_tip}
\end{center}
\end{figure}

\noindent{\bf FIG. \ref{crack_tip}} %18
Comparisons of the positions of the crack-tip as a function of time.
The dash line is the result of the full MD simulation, the solid 
line is the result of the coupled method.
\bigskip

\begin{figure} %19
\begin{center}

\resizebox{4in}{!}{\includegraphics{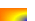}}

\

\resizebox{4in}{!}{\includegraphics{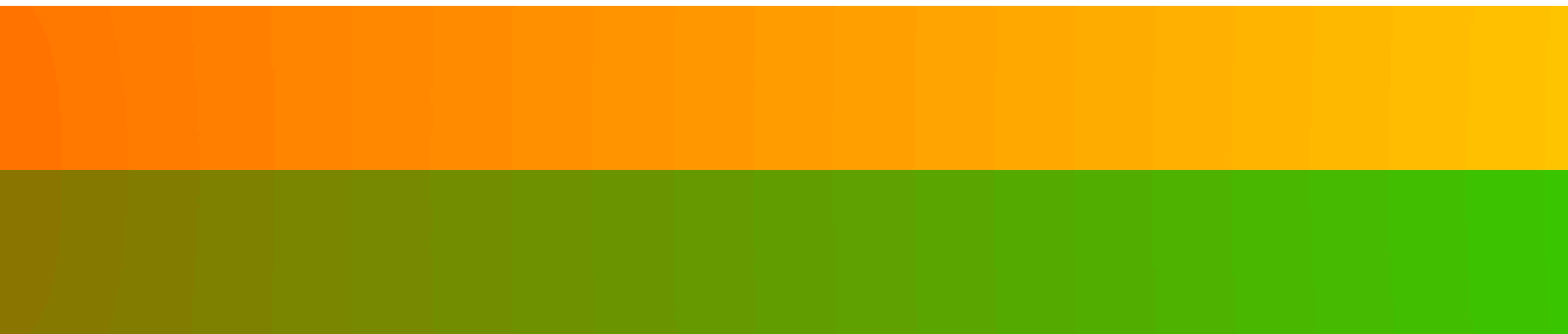}}

\

\resizebox{4in}{!}{\includegraphics{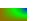}}
\caption{}
\label{shearwave}
\end{center}
\end{figure}

\noindent{\bf FIG. \ref{shearwave}} %19
The shear waves.
We divide the whole domain with $2048\tm 2048$ atoms into three parts. 
The middle part including the crack surface with $2048 \tm 128$ atoms
is the MD region. The top and bottom parts are continuum regions with
$128 \tm 62$ finite difference grids in each region.

\begin{figure} %20
\begin{center}
\resizebox{4in}{!}{\includegraphics{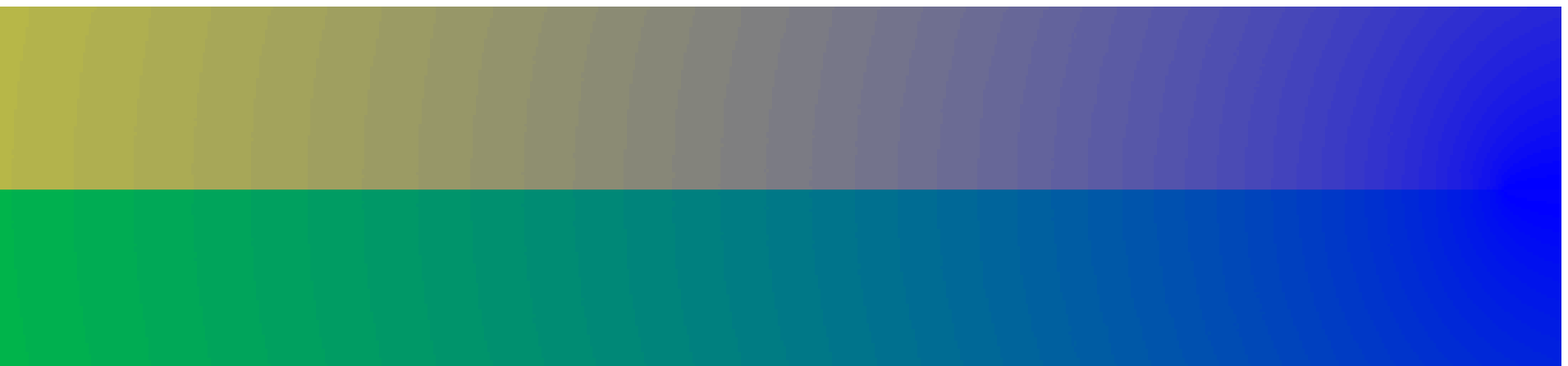}}
\caption{}
\label{shearwave_elarged}
\end{center}
\end{figure}

\noindent{\bf FIG. \ref{shearwave_elarged}} %20
The enlarged picture of the MD region near the crack-tip.
\bigskip

\clearpage 

\clearpage 

\clearpage

\clearpage

\clearpage

\clearpage

\clearpage

\clearpage

\clearpage

\clearpage

\clearpage

\clearpage

\clearpage

\clearpage

\clearpage

\clearpage

\clearpage

\clearpage

\clearpage

\clearpage

\end{document}